\documentclass[aps,showpacs,twocolumn,superscriptaddress]{revtex4}
\usepackage[dvips]{graphicx}
\usepackage{amssymb}
\usepackage{amsmath}
\usepackage{color}

\newcommand{\btab}{\begin{tabbing}}
\newcommand{\etab}{\end{tabbing}}

\newcommand{\beqn}{\begin{equation}}
\newcommand{\eeqn}{\end{equation}}
\newcommand{\barr}[1]{\begin{array}{#1}}
\newcommand{\earr}{\end{array}}
\newcommand{\beqna}{\begin{eqnarray}}
\newcommand{\eeqna}{\end{eqnarray}}
\newcommand{\btablec}{\begin{table} \begin{center}}
\newcommand{\etablec}{\end{center} \end{table}}

\newcommand{\gapproxeq}{\lower.7ex\hbox{$\;\stackrel{\textstyle>}
{\sim}\;$}}
\def\slash#1{#1 \hskip -0.5em / }

\begin{document}

\title{On the near-threshold incoherent $\phi$ photoproduction on the deuteron:\\
any trace of a resonance?}
\author{Alvin Kiswandhi}
\affiliation{Department of Physics and Center for Theoretical Sciences, National Taiwan University,
Taipei 10617, Taiwan}
\affiliation{Department of Physics, STKIP Surya, Tangerang 15810, Indonesia}
\author{Shin Nan Yang}
\affiliation{Department of Physics and Center for Theoretical Sciences, National Taiwan University,
Taipei 10617, Taiwan}
\author{Yu Bing Dong}
\affiliation{Institute of High Energy Physics, Chinese Academy of Sciences, Beijing 100049, China}
\affiliation{Theoretical Physics Center for Science Facilities (TPCSF), CAS, Beijing 100049, China}

\date{\today}

\begin{abstract}

We study the near-threshold incoherent $\phi$ photoproduction on the deuteron based on a model of $\gamma N \to \phi N$, consisting of Pomeron, $(\pi, \eta)$ exchanges,
and a $J^P = 3/2^-$ resonance, which describes the low energy $\gamma p \to \phi p$ LEPS data well, including the peak in the   forward differential cross section.
 The calculation is done up to double rescatterings,  with the spin dependence of the elementary $\gamma N \to \phi N$
amplitude retained throughout the calculation. The Fermi motion and final-state interactions (FSI) are all properly
treated as prescribed by realistic nucleon-nucleon interaction. The couplings of the resonance to $\gamma n$ and $\phi n$ channels are
estimated with the help of a constituent quark model.   The main features of the LEPS and CLAS data are described reasonably well
except for some quantitative  discrepancies  at very low energies and low momentum transfers regions. It is found  that contributions  of Fermi motion, $pn$ FSI, and resonance are all
indispensable in bridging the differences between the single-scattering results and the data.   The off-shell rescattering is found
to be important as it cancels out a large portion of the on-shell contribution.
The discrepancies at  low momentum transfer region  might be related to the binning size of the data. No peak is found to be associated with the  weak resonance as it  gets smeared out by the Fermi motion and FSI with deuterium target. The problem at very low energy region hints at the possible contributions from other
mechanisms and should be investigated in depth with the use of recent high statistics $\gamma p \to \phi p$ data from CLAS.

\end{abstract}


\maketitle

\section{Introduction}
It has long been established that the diffractive processes dominate the $\phi$-meson photoproduction
reaction at high energies and can be well described by $t$-channel Pomeron $(P)$ exchange
\cite{bauer78,donnachie87}. In the low-energy region, the
nondiffractive processes of pseudoscalar $(\pi, \eta)$-meson
exchanges are also known to contribute \cite{bauer78}. Other
processes, such as nucleon exchange \cite{williams98,oh01}, nucleon
resonances \cite{zhao99,titov03}, second Pomeron exchange,
$t$-channel scalar meson and glueball exchanges
\cite{titov99,titov03}, and $s\bar s$-cluster knockout
\cite{titov97a,titov97b,oh01} have also been suggested and studied.
However, no definite conclusion has been inferred because of the
limited experimental data.
Recently, a nonmonotonic behavior in the differential cross sections (DCS) of $\phi$ photoproduction on
proton at forward angles around $E_{\gamma}\sim 2.0$ GeV
has been observed by the LEPS Collaboration \cite{leps05}, and confirmed by the new high statistics data
from CLAS \cite{Dey14,Sera14}. It cannot be
explained by the processes mentioned above.

We found in Refs. \cite{Kiswandhi10,Kiswandhi12} that, with an addition of a resonance
of spin-parity $J^P=3/2^-$, with mass $M = 2.10 \pm 0.03$ GeV and width $\Gamma =
0.465 \pm 0.141$ GeV  to the background mechanisms which consist of
Pomeron and $(\pi, \eta)$-meson exchanges in $t$ channel, not only
the  peak in the forward differential cross section but also the
$t$ dependence of
DCS, $\phi$-meson
decay angular distribution, and the spin-density matrix elements (SDME) can
be well described. It would hence be of interest to see how such a postulated
resonance would exhibit itself in other reactions, like $\phi$-meson photoproduction
from deuterium at low energies.

Data on incoherent photoproduction of $\phi$ meson from deuteron at low
energies have recently become available from LEPS
\cite{Chang10,Chang10DCS} and CLAS \cite{Qian11} Collaborations. While CLAS \cite{Qian11} measured only the
DCS and the decay angular distributions of $\phi$ meson, LEPS provided more extensive data. With a linearly
polarized photon beam, they were able to measure the decay asymmetries and
SDME, in addition
to the
DCS \cite{Chang10,Chang10DCS}.
This prompts us to set forth to
confront our model of Refs. \cite{Kiswandhi10,Kiswandhi12} with these recent extensive data to see whether it is possible to mine this
postulated resonance from them.  This is clearly a daunting task because the strength of this postulated
resonance was found to be relatively weak and can be marred by Fermi motion, final-state interactions (FSI)
of the nucleons, meson rescattering effects, as well as  production via neutron.

The LEPS \cite{Chang10,Chang10DCS} and CLAS \cite{Qian11} data are recently analyzed in Ref. \cite{Sekihara12}, where the Fermi motion
is taken into account for the single-scattering calculation and the effects of the rescatterings of $\phi N$ and $NN$ in the final state
are also investigated. It is found there that both the Fermi motion and the FSI of the nucleons give non-negligible contributions while the double scattering of
the $\phi$ meson with nucleons can be neglected.
 Since we are interested in finding possible trace of the postulated resonance,   which is rather weak, from other reactions like $\phi$-meson photoproduction
from deuteron, a very careful treatment of the reaction is hence essential.

Consequently, in addition to the inclusion of the postulated resonance, we implement the following improvements over the calculation of Ref. \cite{Sekihara12}.
First is that the spin structure of the elementary $\gamma N\rightarrow\phi N$ amplitude, including those of the  Pomeron and $t$-channel ($\pi, \eta$) exchanges, is retained.
 The spin-dependent part of the Pomeron exchange amplitude was extensively studied in Ref. \cite{titov03} and found to be
responsible  for the spin-flip transitions at forward production angles and appears in the angular distributions of $\phi$ decay with both unpolarized and polarized
photon beams. In the present case of incoherent $\phi$ production from deuteron, SDME would either be a constant or zero. The inclusion of the spin-dependent part of the
elementary amplitude will provide useful probe for the resonance and deeper insight into the reaction mechanism.
Secondly, the $t$-channel $(\pi, \eta)$ exchanges are included. This is important because
we found in Ref. \cite{Sekihara12} that the nonmonotonic structure observed in the cross section of $\gamma p \rightarrow \phi p$ is enhanced by the interference of
resonance and $\pi$-exchange contributions. Lastly, the rescattering effects between nucleons are estimated with the realistic Nijmegen $NN$ interactions
\cite{Nijmegen95,Nijmegen15} which ensure that the two-body unitarity is satisfied. Besides, the  $D$ state in the deuteron and the off-shell rescattering are also taken into account.   All of these were overlooked in Ref. \cite{Sekihara12}.

Reactions with deuteron target are often used to extract the corresponding elementary reaction amplitude from neutron. However, in this study, we shall use isospin invariance to infer
the $\phi$ production amplitude from neutron as much as possible. Since it is known that Pomeron behaves like an isoscalar particle, the Pomeron exchange amplitude will
be taken to be  the same as that with proton. For the $t$-channel $(\pi, \eta)$ exchanges, isospin symmetry allows us to write down the corresponding amplitude with neutron.
The only unknown quantity in our model
is just the excitation strength of the resonance from neutron. For our present purpose,
we will take it to be  similar to a resonance with the same spin-parity and roughly the same mass
as well as assuming that it has the same ratio of the proton helicity amplitudes
predicted by a constituent quark model of Ref. \cite{Capstick92}. The details are expounded in Appendix \ref{app:model}.  With this choice,  our results will be
free from  any fitting to the $\gamma d \to \phi p n$
DCS and SDME data obtained by the LEPS and CLAS Collaborations \cite{Chang10DCS,Chang10,Qian11} and some of these results were reported in Ref. \cite{Yang14}.

This paper is organized as follows. In Sec. \ref{sec:model} we   present the details of our calculations. The elementary $\gamma N\rightarrow \phi N$ amplitude is  first briefly described. Then the $\phi$ production amplitudes via single and double scattering mechanism with deuteron
target are given. In Sec. \ref{sec:results}, results are shown
and discussed. Summary and conclusions are presented in Sec. \ref{sec:conclusions}. Some details of our calculations are given in the appendices for clarity.

\section{The model for $\gamma d \to \phi p n$ reaction}
\label{sec:model}
In this section, we present the essentials of our calculations. The kinematics and the notations are first introduced. Then the elementary
amplitude for photoproduction of a $\phi$ meson from nucleon, $\gamma  N \to \phi N$, the basis of our calculation, is briefly discussed
with details given in Appendix A. Lastly, we explain the details of our calculations, regarding how the Fermi motion and final state rescattering with both
on- and off-shell ones, are treated.

\subsection{Kinematics}
\label{sec:kinematics}

\begin{figure}[h]

\begin{center}
\includegraphics[width=0.35\linewidth,angle=0]{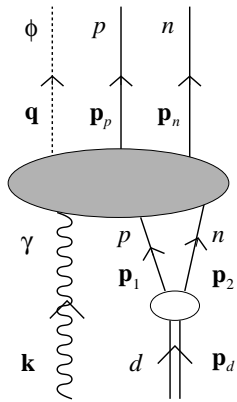}
\end{center}

\caption{{The $\gamma d \to \phi p n$ reaction with $\gamma ({\mathbf{k}})$, $\phi ({\mathbf{q)}}$, $d ({\mathbf{p}}_d)$, $p ({\mathbf{p}}_p)$, and $n ({\mathbf{p}}_n)$ denote the photon, $\phi$ meson, deuteron, proton, and neutron lines, respectively, with their momenta given inside the brackets. Also, ${\mathbf{p}}_1$ (${\mathbf{p}}_2$) denotes the initial proton (neutron) momenta inside the deuteron. It should be emphasized that the ellipse joining the deuteron, proton, and neutron lines is not an interaction vertex.}}
\label{general}
\end{figure}

Let us first introduce the momenta of the particles involved in the reaction. Here, $k$, $p_d$, $q$, $p_p$, and $p_n$ are the four-momenta of the
photon, deuteron, $\phi$ meson, proton, and neutron, respectively, while $p_1$ ($p_2$) is that of the proton (neutron) inside the deuteron, as shown in Fig. \ref{general}.
Notice also that $k = (E_\gamma, {\mathbf{k}})$, $q = (E_\phi, {\mathbf{q}})$, and $p_a = (E_a,{\mathbf{p}}_a)$ where $a = p, n, d$.
The masses of the deuteron, $\phi$ meson, proton, and neutron are denoted by $M_d$, $M_\phi$, $M_p$, and $M_n$, respectively. We work in the laboratory (LAB) frame where the deuteron is
at rest. In this study, we use  the plane wave normalization that reads $\langle {\mathbf{p}}' | {\mathbf{p}} \rangle = (2\pi)^3 2 E_p \delta^{(3)}({\bf p}' - {\bf p})$ and $\bar{u}({\bf{p}},s)u({\bf{p}},s) = 2M$ for a Dirac spinor
with mass $M$. In addition, we introduce
\beqn
\langle f |\hat T| i \rangle= (2 \pi)^4 \delta^{(4)} \left(k + p_d - q - p_p - p_n\right) {\mathcal{M}}_{fi},
\eeqn
where the $S$-matrix is given by $\hat S = \hat I - i\hat T$.
The invariant amplitude $-i{\mathcal{M}}_{fi}$ is obtained diagrammatically with Feynman rules.

For later convenience, we define the Mandelstam variables $s$, $t_\phi$, and $u_\phi$ as follows,
\beqna
s &=& (k + p_d)^2 = (q + p_p + p_n)^2, \nonumber \\
t_\phi &=& (q - k)^2 = (p_d - p_p - p_n)^2, \nonumber \\
u_\phi &=& (q - p_d)^2 = (k - p_p - p_n)^2,
\eeqna
and
\beqn
s + t_\phi + u_\phi = M_d^2 + M_\phi^2 + M_{pn}^2,
\eeqn
where $M_{pn}$ is the invariant mass of the $pn$ system in the final state. For a fixed value of $t_\phi$, $M_{pn}$ has the minimum and maximum values of,
\beqna
M_{pn}^{\textrm{min}} &=& M_p + M_n, \nonumber \\
M_{pn}^{\textrm{max}} &=& \sqrt{s + M_\phi^2 - 2 \sqrt{s} E_\phi'},
\eeqna
with
\beqna
E_\phi' &\equiv& \sqrt{M_\phi^2 + q'^2}, \nonumber \\
q' &\equiv& \frac{M_\phi^2 - t_\phi}{4 E_\gamma^{\textrm{cm}}} - \frac{M_\phi^2 E_\gamma^{\textrm{cm}}}{M_\phi^2 - t_\phi}.
\eeqna
The value of $u_\phi$ is in turn  limited  to be within
\beqna
u_{\phi, \textrm{min}} &=& M_d^2 + M_\phi^2 + (M_p + M_n)^2 - s - t_\phi, \nonumber \\
u_{\phi, \textrm{max}} &=& M_d^2 + 2 M_\phi^2 - 2 \sqrt{s}E_\phi' - t_\phi,
\eeqna
and the value of $t_\phi$ is also restricted within
\beqna
t_{\phi, \textrm{min}} &=& M_\phi^2 - 2 E_\gamma^{\textrm{cm}}(E_{\phi, \textrm{max}} - q_{\textrm{max}}), \nonumber \\
t_{\phi, \textrm{max}} &=& M_\phi^2 - 2 E_\gamma^{\textrm{cm}}(E_{\phi, \textrm{max}} + q_{\textrm{max}}),
\eeqna
where  $E_\gamma^{\textrm{cm}}$ is the photon energy in the $\gamma d$ center-of-mass (CM) frame and
\beqna
q_{\textrm{max}} &\equiv& \sqrt{\frac{[s - (M_\phi + M_{pn}^{\textrm{min}})^2][s - (M_\phi - M_{pn}^{\textrm{min}})^2]}{4s}}, \nonumber \\
E_{\phi, \textrm{max}} &\equiv& \sqrt{M_\phi^2 + q_{\textrm{max}}^2},
\eeqna
which corresponds to the case where the three-momentum of the $\phi$ meson in the CM system achieves its maximum value which clearly happens only when
$M_{pn}$ is in its minimum.

The differential cross section of $\gamma d \to \phi p n$ in the LAB system is
\beqna
\frac{d \sigma_d}{dt_\phi} &=& \frac{1}{128 E_\gamma^2 M_d^2} \frac{1}{(2 \pi)^4} \int_{u_{\phi, \textrm{min}}}^{u_{\phi, \textrm{max}}} d u_\phi \frac{p^{c,pn}_{p}}{M_{pn}}
\nonumber \\
&\times& \int d\Omega^{c,pn}_{p} \bar{\sum_\lambda} \sum_{\lambda'} \left|{\mathcal{M}}_{fi}\right|^2,
\eeqna
where $\lambda$ ($\lambda'$) denotes the initial (final) spins, and  ${\mathbf{p}}^{c,pn}_{p} = (p^{c,pn}_{p}, \Omega^{c,pn}_{p})$ denotes the three-momentum of the final proton in
the CM system of the final $pn$ system.

\subsection{The elementary $\gamma N \to \phi N$ amplitudes}
\label{sec:gamma_N}

The basic input in our model  is the elementary amplitude of
$\phi$-meson photoproduction from a free nucleon,
${\mathcal{M}}_{\gamma N \to \phi N}$ in which $N = p ,n$.
In our study, the amplitude ${\mathcal{M}}_{\gamma p \to \phi p}$ constructed in our previous work \cite{Kiswandhi12}.
will be employed. It consists of nonresonant and resonant amplitudes. The nonresonant amplitude consists of Pomeron and $(\pi,\eta)$ exchanges in the
$t$ channel. The resonant amplitude arises from a postulated $J^P = 3/2^-$ resonance contribution. Notice  that the contribution from the $u$-channel amplitude is very small
and in this work, we will include only the $s$-channel contribution. The details of the model are given
in Appendix \ref{app:model}.  The values for the mass, width, and coupling constants for the $J^P = 3/2^-$ resonance, as determined in Ref. \cite{Kiswandhi12}, are presented in Table. \ref{tab:nstar}.

For the production amplitude from neutron ${\mathcal{M}}_{\gamma n \to \phi n}$, the $P$ and $(\pi,\eta)$ $t$-channel exchanges amplitudes
can be readily written down with the assumption of isospin symmetry.  However,  the resonance couplings to $\gamma n$ and $\phi n$ channels have yet to be determined
as there is no data on $ \gamma n \to \phi n$ available. For our present purpose,  we will determine their values according to the following
recipe. Namely,   we first assume that the electromagnetic excitation of the resonance, hence its ratio of helicity amplitudes $A_{1/2}^p/A_{3/2}^p$ for $\gamma p$,
would be similar to that of a $J^P = 3/2^-$ nucleon state with roughly the same mass as predicted by a theoretical model.
We will take it as the $J^P = 3/2^-$ nucleon state with a bare mass of 2095 MeV and a positive value for the ratio of helicity amplitudes
$A^p_{1/2}/A^p_{3/2}$ for $\gamma p$ decay as predicted in the constituent quark model (CQM) of Ref. \cite{Capstick92}. For the resonance coupling with $\phi n$ channel,
we assume that they are identical to that with the $\phi p$ channel since $\phi$ is an isoscalar particle.
The details of the  determination of the coupling constants
$g_{\gamma n n^*}^{(i)}$ and $g_{\phi n n^*}^{(j)}$ of $\gamma n n^*$ and $\phi n n^*$ vertices are presented in Appendix \ref{app:model}.

 Note that the Pomeron-exchange amplitude as given in Eq. (\ref{ampa}) depends on the polarizations of both the incident photon and outgoing $\phi$ meson
which were neglected in Ref. \cite{Sekihara12}. The $\pi$- and $\eta$-exchange amplitudes were also not included in Ref. \cite{Sekihara12}.

\begin{table}
\caption{\label{table} The $N^*$ mass, width, and coupling constants for $J^P = 3/2^-$ resonances.}

\begin{center}
\begin{tabular}{lcc}
\hline\hline
                                       & Proton $(N = p)$     & Neutron $(N = n)$ \\
\hline

$M_{N^*}$(GeV)                         & \multicolumn{2}{c}{2.08} \\
$\Gamma_{N^*}$(GeV)                    & \multicolumn{2}{c}{0.570} \\

\hline

${g}_{\gamma N N^*}^{(1)}$             &  $\quad \qquad 0.0323 \qquad \quad$        & $\quad \qquad -0.0441 \qquad \quad$  \\

${g}_{\gamma N N^*}^{(2)}$             &  $0.0420$        & $-0.0193$  \\

\hline

${g}_{\phi N N^*}^{(1)}$               &  \multicolumn{2}{c}{$-20.94$} \\

${g}_{\phi N N^*}^{(2)}$               &  \multicolumn{2}{c}{$-2.61$}   \\

${g}_{\phi N N^*}^{(3)}$               &  \multicolumn{2}{c}{$-3.36$}   \\

\hline\hline
\end{tabular}
\end{center} \label{tab:nstar}
\end{table}

\subsection{$\gamma d \to \phi p n$ amplitudes}
\label{sec:gamma_d}

\begin{figure*}[htbp]

\begin{center}
\includegraphics[width=0.90\linewidth,angle=0]{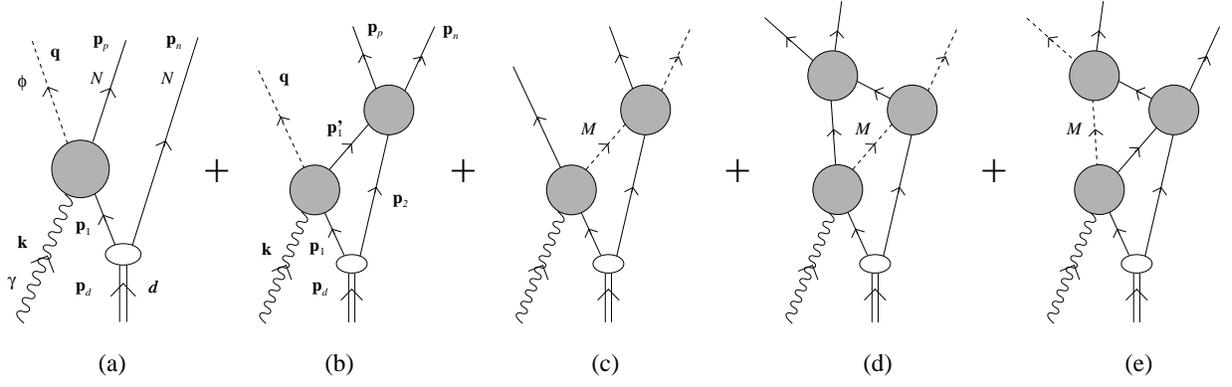}
\end{center}

\caption{The diagrams of $\gamma d \to \phi p n$ reaction up to triple scatterings.  Here, meson $M = \pi, \eta, \rho, \omega, \phi, \cdots$.  (a) is the single-scattering diagram, (b) double-scattering diagram with FSI between the final nucleons, (c) double-scattering diagram with FSI between meson and final nucleon, (d) and (e) triple-scattering diagrams. In this work we consider only the contributions from the diagrams (a)-(c), namely, up to
 double scatterings.}
\label{all_diagram}
\end{figure*}

 Within the multiple scattering scheme,  the diagrams for $\gamma d \to \phi p n$ reaction up to triple rescatterings are shown in Fig. \ref{all_diagram}.
Fig. \ref{all_diagram}(a) is the single-scattering diagram, \ref{all_diagram}(b) double-scattering diagram with rescatterings between the final nucleons, and
\ref{all_diagram}(c) the double-scattering diagram with rescatterings between meson $M$ produced
by the incoming photon and final nucleon. Figs. \ref{all_diagram}(d) and (e) represent the triple-scattering diagrams.
In this work, we will consider only up to the single- and $pn$ double-scattering diagrams  of Figs. \ref{all_diagram}(a) and (b),
as Fig. \ref{all_diagram}(c) with $M = \pi, \eta, \rho, \omega, \phi, \cdots$ was studied in Ref. \cite{Sekihara12} and found to be small.

 In general, the amplitude for $\gamma d \to \phi p n$ can be expressed as
\beqna
&&{\mathcal{M}}_{fi} = \sum_{m_1, m_2}\int \frac{d^3 p_1}{(2 \pi)^3} \frac{d^3 p_2}{(2 \pi)^3}\frac{1}{2 E_1}\frac{1}{2 E_2} \nonumber \\
&& \times {\mathcal{M}}_{\gamma p n}({\mathbf{k}}, {\mathbf{p}}_1, {\mathbf{p}}_2, m_\gamma, m_1, m_2; {\mathbf{q}}, {\mathbf{p}}_p, {\mathbf{p}}_n, m_\phi, m_p, m_n)
\nonumber \\
&& \times\langle{\mathbf{p}}_1, {\mathbf{p}}_2; m_1, m_2|{\mathbf{p}}_d, \Phi_d; m_d\rangle,
\label{general_eq}
\eeqna
where the momenta of the particles are defined in Fig. \ref{general} and the internal structure of the deuteron is characterized by its wave function $\Phi_d$.
The spin projections of the particles are denoted by $m$ and the naming follows that of the momenta.
Since the Pomeron amplitude contains spin-spin and spin-orbital
dependent terms which are responsible for the spin-flip transition at forward angles and affect  the angular distribution of $\phi \rightarrow K^+ K^-$, we will  include the 
$D$ state of the deuteron in our calculation.

It is important to note that Eq. (\ref{general_eq}) explicitly implies that deuteron is treated nonrelativistically in our study. It has the consequence that the intermediate nucleons with three-momentum
${\mathbf{p}}_1$ and ${\mathbf{p}}_2$ in Fig. \ref{general} are both on mass-shell and the energy in the intermediate states is not conserved, namely, $E_1({\mathbf{p}}_1) + E_2({\mathbf{p}}_2) \neq E_d$,
where $E_i({\mathbf{p}}_i)=(M^2+{\mathbf{p}}_i^2)^{1/2}$. This point will be brought up often in the subsequent discussion.

 In this study, the deuteron wave function $\Phi_d$, including that of $D$ state, as prescribed from the Bonn potential \cite{Machleidt01} will be employed.

\subsubsection{Single-scattering amplitudes}
\label{sec:gamma_d_single}

The diagram for single-scattering amplitude from a deuteron target with no FSI is shown in Fig. \ref{all_diagram}(a).
 We have, for the amplitude in which the $\phi$ meson is produced from one of the nucleons
in the deuteron, say,  proton,
\beqna
&&{\mathcal{M}}_{\gamma p n}^{(s,p)}({\mathbf{k}}, {\mathbf{p}}_1, {\mathbf{p}}_2, m_\gamma, m_1, m_2; {\mathbf{q}}, {\mathbf{p}}_p, {\mathbf{p}}_n, m_\phi, m_p, m_n) \nonumber \\
&=& (2 \pi)^3 2 E_2\delta^{(3)}\left({\mathbf{p}}_n - {\mathbf{p}}_2\right) \delta_{m_n m_2} \nonumber \\
&& \times   {\mathcal{M}}_{\gamma \phi} \left(\mathbf{k}, \mathbf{p}_1, m_\gamma, m_1; \mathbf{q}, \mathbf{p}_p, m_\phi, m_p\right),
\label{singscatt p}
\eeqna
where the superscripts $s$  and $p$ refer to the fact that the amplitude arises from single scattering in which the $\phi$ meson
is produced on a proton target.

  The amplitude ${\mathcal{M}}_{\gamma \phi}$  in Eq. (\ref{singscatt p}) denotes the $\gamma p \to \phi p$ elementary amplitude with the spins and momenta of the
particles specified within the parentheses. It should be noted that this is not the  one obtained from the Feynman rules in which the four-momenta are conserved. This is because we treat  deuteron nonrelativistically so that the struck proton is on mass-shell.  Accordingly, only the three-momentum in the subprocess $\gamma p \to \phi p$ would be conserved but not the energy, namely, the energies of
the initial state $E_\gamma(\mathbf{k})+E_p(\mathbf{p}_1)$ and the final state $E_\phi(\mathbf{q})+ E_p(\mathbf{p}_p)$ do not have to be equal.

In this work, we will adopt the following recipe to obtain ${\mathcal{M}}_{\gamma \phi}$  in Eq. (\ref{singscatt p}).
We start from the corresponding invariant amplitudes ${\mathcal{M}}_{\gamma \phi}$ with $t=(q-k)^2$ and $s=(q+p_p)^2$ as given in Eqs. (\ref{ampa},\,\,\ref{ampb}) for the Pomeron and $(\pi, \eta)$ exchange amplitudes ${\mathcal{M}}_P$
and ${\mathcal{M}}_{\pi +\eta}$ of Appendix \ref{app:model} as well as the resonance amplitude  outlined there and then express them in terms of the respective momenta and spin variables of all four particles of $\gamma, \phi$ and the incoming and outgoing nucleons in the CM frame.
We first transform the four-momenta
of the particles in the initial (final) state to the CM frames of the initial (final) state. Here, we have
$E_\gamma^{CM}(\mathbf{p}_i^{CM})+E_p^{CM}(-\mathbf{p}_i^{CM})$ and
$E_\phi^{CM}(\mathbf{p}_f^{CM})+ E_p^{CM}(-\mathbf{p}_f^{CM})$ as the initial and final total energies in the CM frame, respectively.
We then take an approximation where, using the notation $\mathbf{p}_i^{CM} = (|\mathbf{p}_i^{CM}|, \Omega_i^{CM})$,
the initial momentum magnitude $|\mathbf{p}_i^{CM}|$ is
chosen such that the resulting initial total energy would be equal to that of the final. Notice that the direction of the momentum $\Omega_i^{CM}$ is kept constant.
It is reasonable to do this since the actual energy of the system is the energy of $\phi N$ in the final state. The spins of the particles are also in general different after the Lorentz transformation. They are transformed by using the proper transformation
for each spin. The photon and $\phi$ meson polarization wave functions are transformed by using the ordinary Lorentz transformation for four momenta, while the spinor of the
nucleon is transformed by using the Lorentz transformation for spin-1/2 spinor.

 To obtain the contribution of the $\phi$ production produced from proton in the deuteron via single scattering to $\mathcal{M}_{fi}$, the amplitude ${\mathcal{M}}_{\gamma p n}^{(s,p)}$ of Eq. (\ref{singscatt p}) should be convoluted with the deuteron
wave function as  in Eq. (\ref{general_eq}).

The amplitude in which the $\phi$ meson is produced on the neutron can be written down similarly as
Eq. (\ref{singscatt p}) with some suitable changes.

\subsubsection{Double-scattering amplitudes}
\label{sec:gamma_d_double}

The double-scattering diagram in which the outgoing proton and neutron rescatter  is shown in Fig. \ref{all_diagram}(b) and the corresponding amplitude
can be expressed as, neglecting the spins
\beqna
&&{\mathcal{M}}_{\gamma p n}^{(d)}({\mathbf{k}}, {\mathbf{p}}_1, {\mathbf{p}}_2; {\mathbf{q}}, {\mathbf{p}}_p, {\mathbf{p}}_n) \nonumber \\
&=&\bar{u}({\mathbf{p}}_p)\bar{u}({\mathbf{p}}_n)\hat{M}_{NN}u({\mathbf{p}}_2){(\slash{p}_1' + m) \over p_1'^2 - m^2 + i\epsilon} \nonumber \\
&& \times \epsilon^*_\mu({\mathbf{q}})\hat{M}_{\gamma \phi}^{\mu\nu}u({\mathbf{p}}_1)\epsilon_\nu({\mathbf{k}}),
\label{double_scatt}
\eeqna
in which the superscript $d$ denotes that the amplitude arises from double-scattering process. Here, $\hat{M}_{NN}$ and $\hat{M}_{\gamma \phi}^{\mu\nu}$ are related to the invariant amplitude ${\mathcal{M}}$ by the relations
\beqn
{\mathcal{M}}_{NN} = \bar{u}({\mathbf{p}}_3)\bar{u}({\mathbf{p}}_4)\hat{M}_{NN}u({\mathbf{p}}_1)u({\mathbf{p}}_2),
\eeqn
for the reaction $N({\mathbf{p}}_1) + N({\mathbf{p}}_2) \to N({\mathbf{p}}_3) + N({\mathbf{p}}_4)$, and
\beqn
{\mathcal{M}}_{\gamma \phi} = \epsilon^*_\mu({\mathbf{q}})\bar{u}({\mathbf{p}}')\hat{M}_{\gamma \phi}^{\mu\nu}u({\mathbf{p}})\epsilon_\nu({\mathbf{k}}),
\eeqn
for the reaction $\gamma({\mathbf{k}}) + N({\mathbf{p}}) \to \phi({\mathbf{q}}) + N({\mathbf{p}}')$.

As mentioned earlier, the nucleons in the deuteron with three-momentum $\mathbf{p}_i$ in Fig. \ref{all_diagram}(b) will be treated as on-mass-shell and
hence would propagate only forwardly. So, only the positive-energy component in the Feynman propagator
\beqn
S_F(p_1')={(\slash{p}_1' + m) \over p_1'^2 - m^2 + i\epsilon}
\eeqn
for the nucleon intermediate states with four-momentum $p_1'$ in Eq. (\ref{double_scatt})
would be kept. It is then a straightforward exercise to arrive at the following expression
\beqna
&&{\mathcal{M}}_{\gamma p n}^{(d,+)}({\mathbf{k}}, {\mathbf{p}}_1, {\mathbf{p}}_2; {\mathbf{q}}, {\mathbf{p}}_p, {\mathbf{p}}_n) \nonumber \\
&=&{1 \over 2 E_1'}{ {\mathcal{M}}_{NN}(NN \to NN){\mathcal{M}}_{\gamma \phi}(\gamma N \to \phi N)\over E - (E_\phi + E_1' + E_2) + i\epsilon},
\label{double_scattPE}
 \eeqna
where superscript $+$ denotes that only positive energy intermediate states are retained and, ${\mathbf{p}}_1' = {\mathbf{k}} + {\mathbf{p}}_1 - {\mathbf{q}}, E_1'=(M^2+\mathbf{p}^2_1)^{1/2}$, and $E =   E_\phi + E_p + E_n$.
It has to be noticed that  in Eq. (\ref{double_scattPE}), summation over intermediate spins is understood.

 Again   both $\mathcal{M}_{NN}(NN \to NN)$ and $\mathcal{M}_{\gamma \phi}(\gamma N \to \phi N)$ in the above Eq. (\ref{double_scattPE}), only the three-momentum is conserved but not necessarily the energy. We already encounter
this problem for $\mathcal{M}_{\gamma \phi}(\gamma N \to \phi N)$ when discussing the case of single scattering and the same recipe will be followed. For $\mathcal{M}_{NN}(NN \to NN)$, we note that it is what
is called the $t$-matrix element in the potential scattering and given by $t=v+vg_0t$. In addition,   the propagator in Eq. (\ref{double_scattPE}) contains two parts,
\beqna
\frac{1}{E - (E_\phi + E_1' + E_2) + i\epsilon} &=& \mathcal{P}\frac{1}{E - (E_\phi + E_1' + E_2)} \nonumber \\
&-& i\pi\delta[E - (E_\phi + E_1' + E_2)]. \nonumber \\
\label{on-off-shell}
\eeqna
The first and second terms on the right-hand side would correspond to  the half-off- and on-energy-shell rescatterings between the final $pn$ state,  respectively. The
half-off-shell and the on-shell ${\mathcal{M}}_{NN}(NN \to NN)$ matrix elements are hence needed. We evaluate them with the Nijmegen potential \cite{Nijmegen95,Nijmegen15}.
Eq. (\ref{double_scattPE}) has to be convoluted with the deuteron wave function as the case of single scattering.

Another type of double rescattering diagrams involve intermediate mesons like $\pi$, $\eta$, $\rho$, $\omega$, and $\phi$ first produced by the photon
as depicted in Fig. \ref{all_diagram}(c), can also be treated in the similar manner as Fig. \ref{all_diagram}(b) as outlined in the above. However, it can be estimated to be
small as follows.
 We  realize that the total cross sections $\sigma(i)$ arising from an intermediate state $i$ are
roughly proportional to the product of the cross sections of its intermediate reactions
\beqna
&&\sigma(NN) \propto \sigma_{\gamma N \to \phi N}\sigma_{NN \to NN}\nonumber \\
&&\sigma(\phi N) \propto \sigma_{\gamma N \to \phi N}\sigma_{\phi N \to \phi N}\nonumber \\
&&\sigma(\pi N) \propto \sigma_{\gamma N \to \pi N}\sigma_{\pi N \to \phi N}.
\eeqna

\begin{figure*}[htbp]
\begin{center}
\includegraphics[width=0.75\linewidth,angle=0]{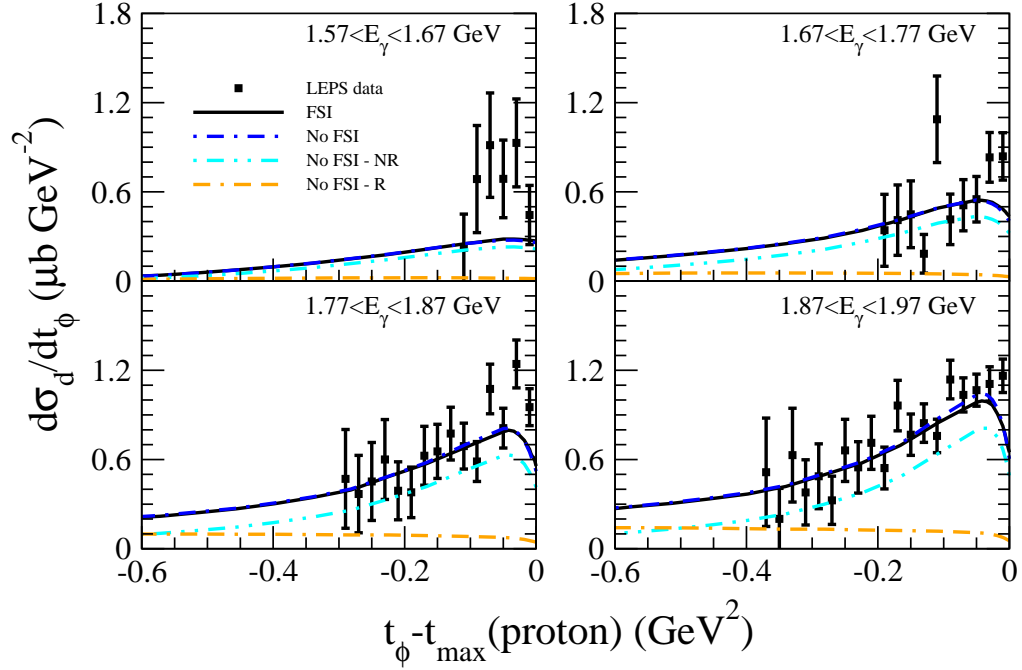}
\end{center}

\vspace{-1.0cm}

\caption{The DCS of $\gamma d \to \phi p n$ as a function of $t_\phi$ at four energy bins $1.57 < E_\gamma < 1.67$ GeV, $1.67 < E_\gamma < 1.77$ GeV,
$1.77 < E_\gamma < 1.87$ GeV, and $1.87 < E_\gamma < 1.97$ GeV. The full lines are the results of $NR + R$ with $pn$ FSI. The dash-dotted, dash-dot-dotted, and
dash-dash-dotted lines are the results of $NR + R$, $NR$, and $R$ without $pn$ FSI, respectively.
Here, $NR$ and $R$ denote nonresonant and resonant amplitudes, respectively. The squares with error bars are the experimental data of Refs. \cite{Chang10DCS, Chang_pc}.}
\label{DCS_1}
\end{figure*}

\begin{figure*}[htbp]
\begin{center}
\includegraphics[width=0.75\linewidth,angle=0]{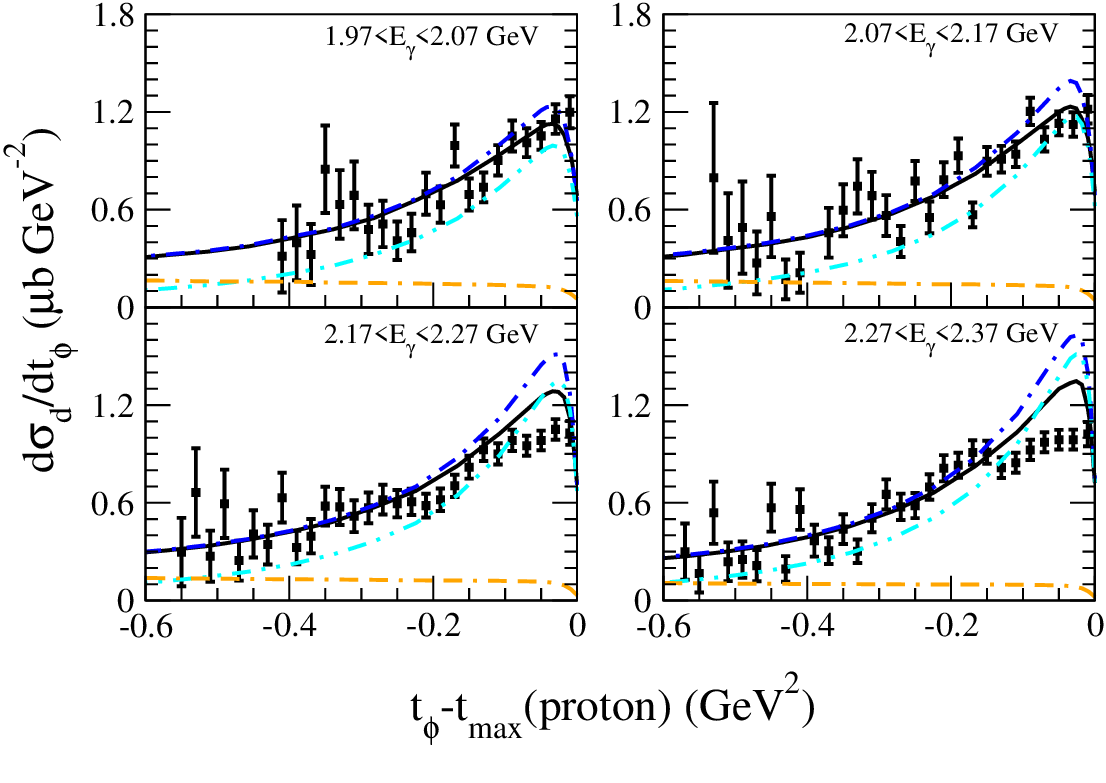}
\end{center}

\vspace{-1.0cm}

\caption{Caption is the same as in Fig. \ref{DCS_1}, but for four energy bins $1.97 < E_\gamma < 2.07$ GeV, $2.07 < E_\gamma < 2.17$ GeV,
$2.17 < E_\gamma < 2.27$ GeV, and $2.27 < E_\gamma < 2.37$ GeV.}
\label{DCS_2}
\end{figure*}

Now, the values for the total cross sections of the intermediate reactions relevant to the kinematic region with photon LAB energy $E_\gamma \sim 2$ GeV
are \cite{Sibirtsev06, Xie08},
\beqna
&&\sigma_{\gamma N \to \phi N} \approx 0.3 \mu b,\nonumber \\
&&\sigma_{\gamma N \to \pi N} \approx 5 \mu b,\nonumber \\
&&\sigma_{NN \to NN} \approx 10^6 \mu b,\nonumber \\
&&\sigma_{\phi N \to \phi N} \approx 11 \cdot 10^3 \mu b,\nonumber \\
&&\sigma_{\pi N \to \phi N} \approx 30 \mu b,
\eeqna
which give
\beqna
&&\sigma(NN) \propto 3 \cdot 10^5,\nonumber \\
&&\sigma(\phi N) \propto 3.3 \cdot 10^3,\nonumber \\
&&\sigma(\pi N) \propto 1.5 \cdot 10^2,
\eeqna
or
\beqn
\sigma(NN):\sigma(\phi N):\sigma(\pi N) = 1 : 1.1 \cdot 10^{-2} : 5 \cdot 10^{-4}.
\eeqn
For the cross sections whose values are not available like $\sigma(\rho N)$ and $\sigma(\omega N)$, it is likely that their values are of the same order of magnitude as
that of $\sigma(\pi N)$, as $\pi$, $\rho$, and $\omega$ are all meson with zero strangeness. For $\sigma(\eta N)$, its value is probably close to that of $\sigma(\phi N)$, as
$\eta$, like $\phi$, contains $s\bar{s}$. Since $\sigma(NN)$ is roughly the same order of magnitude as the DCS arising from
single-scattering process, $\sigma(\phi N)$ and $\sigma(\eta N)$ are then around few percents of it. It can then be concluded that if the $\phi N$ and $\eta N$ intermediate states are
incorporated, at most their contributions are just a few percents of the total cross section. This result is also supported by Ref. \cite{Sekihara12}, where it is shown that the effect arising from the $\phi N$ FSI is small.

\section{Results}
\label{sec:results}

With the model presented  in the last section, it is straightforward to calculate the DCS, SDME, and other  observables of the $\gamma d \to \phi p n$ reaction.
We remind the readers that once our model for the elementary amplitude of $\gamma p \to \phi p $ is fixed and the $\gamma NN^*$ couplings determined as
explained briefly in Sec. \ref{sec:gamma_N} and in details in Appendix \ref{app:model}, our results for $\gamma d \to \phi p n$ are simply predictions and free from any fitting.

 We will  present the results and focus on the role played by the resonances, the effects of Fermi motion, and the FSI before comparing them with the existing data, first for the DCS and then for the SDME.
In our discussion of the SDME, the connections between the spin dependence and the elementary amplitudes, as well as the values of SDME will be pointed out.

\subsection{Differential cross sections}
\label{sec:DCS}

\begin{figure*}[htbp]
\begin{center}
\includegraphics[width=0.75\linewidth,angle=0]{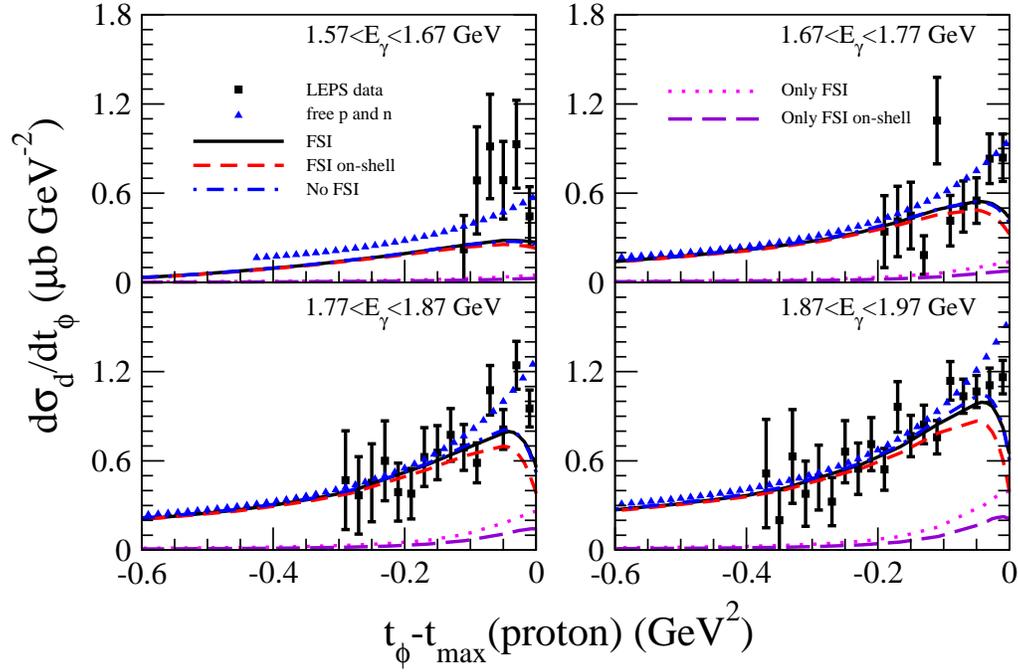}
\end{center}

\vspace{-1.0cm}

\caption{The DCS of $\gamma d \to \phi p n$ as a function of $t_\phi$ at four energy bins $1.57 < E_\gamma < 1.67$ GeV, $1.67 < E_\gamma < 1.77$ GeV,
$1.77 < E_\gamma < 1.87$ GeV, and $1.87 < E_\gamma < 1.97$ GeV. The triangles are the results for the simple summation of free proton and neutron DCS.
The full, dashed, and dash-dotted lines are the results of $NR + R$ with $pn$ FSI, with only on-shell $pn$ FSI, and  without $pn$ FSI, respectively.
The dotted and long-dashed lines are the results of $NR + R$ with $pn$ FSI and with only on-shell $pn$ FSI, respectively, without the single-scattering contribution.
Here, $NR$ and $R$ denote nonresonant and resonant amplitudes, respectively. The squares with error bars are the experimental data of Refs. \cite{Chang10DCS, Chang_pc}.}
\label{DCS_1_medium}
\end{figure*}

\begin{figure*}[htbp]
\begin{center}
\includegraphics[width=0.75\linewidth,angle=0]{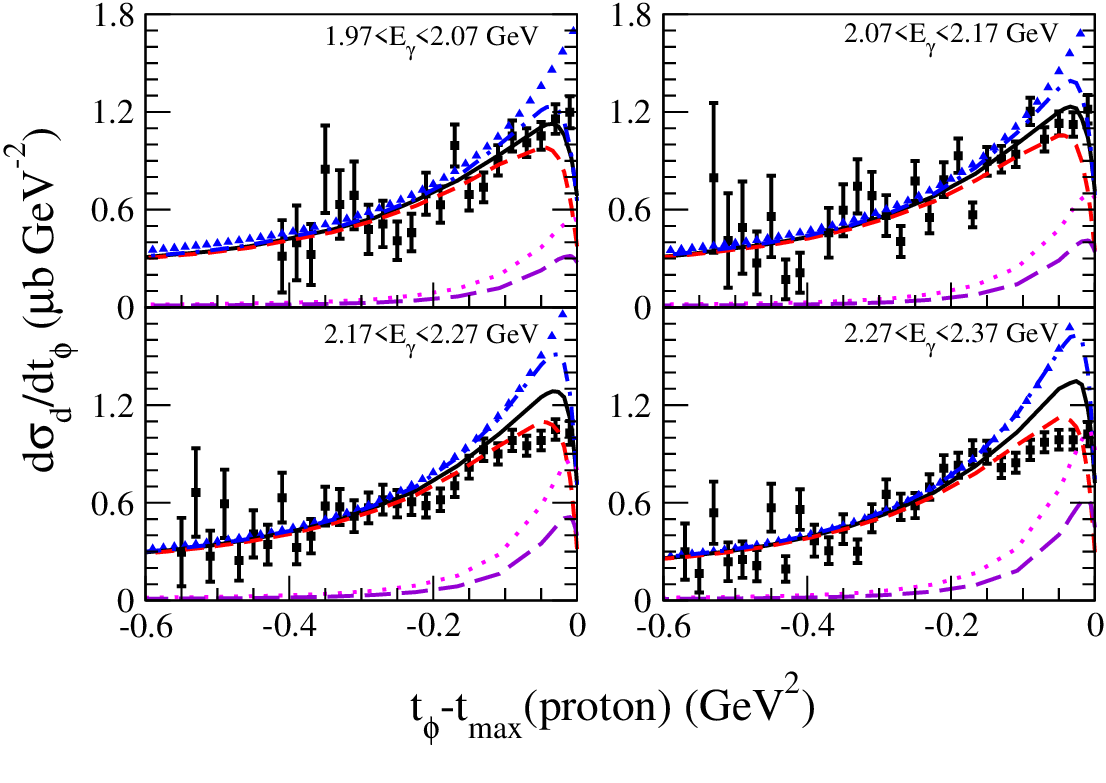}
\end{center}

\vspace{-1.0cm}

\caption{Caption is the same as in Fig. \ref{DCS_1_medium}, but for four energy bins $1.97 < E_\gamma < 2.07$ GeV, $2.07 < E_\gamma < 2.17$ GeV,
$2.17 < E_\gamma < 2.27$ GeV, and $2.27 < E_\gamma < 2.37$ GeV.}
\label{DCS_2_medium}
\end{figure*}

 The DCS as a function of $t_\phi$ at eight energy bins from LEPS \cite{Chang10DCS, Chang_pc} are given in Figs. \ref{DCS_1}, \ref{DCS_2}, \ref{DCS_1_medium}, and \ref{DCS_2_medium}.
The DCS as a function of $t_\phi$ at
$1.65 < E_\gamma < 1.75$ GeV measured by CLAS \cite{Qian11} are shown in Fig. \ref{DCS_CLAS}. The ratio of the DCS with only FSI to the DCS without the FSI at several 
incoming photon LAB energies is given in Fig. \ref{DCS_ratio_1}.
The DCS at $t_\phi = t_{\textrm{max}}$ and their ratio to twice of the production from free proton, where $t_{\textrm{max}}$ corresponds to the value of the maximum $t$ for
free proton case \cite{Chang10DCS, Chang_pc} are given in Fig. \ref{DCS_forward}. The ratio of the DCS from the proton inside the deuteron to that of the free proton case
as a function of photon LAB energy \cite{Chang10DCS, Chang_pc} are given in Fig. \ref{DCS_forward_1}(a).

 The full and dash-dotted lines are the results obtained by including nonresonant and resonant amplitudes with and without $pn$ FSI, respectively. Here, including $pn$ FSI
means that we include both the on- and off-shell rescattering effects. The dashed lines are the results obtained by including
only the on-shell part of the $pn$ FSI. The dash-dot-dotted and dash-dash-dotted lines are the results without $pn$ FSI calculated by including only nonresonant and resonant
amplitudes, respectively. The dotted lines are the results of nonresonant plus resonant amplitudes with $pn$ FSI without the single-scattering contribution.
The long-dashed lines are plotted similarly to the dotted ones, but with only the on-shell part included. Our results are plotted by using the middle values of each bin
and the maximum value of $t$ for the free proton case $t_{\textrm{max}}$(proton) corresponds to these values as well. The squares with error bars are the LEPS data
\cite{Chang10DCS, Chang10, Chang_pc}.

\subsubsection{The role of the resonance}
\label{sec:DCS_res}

\begin{figure}[htbp]

\begin{center}
\includegraphics[width=0.8\linewidth,angle=0]{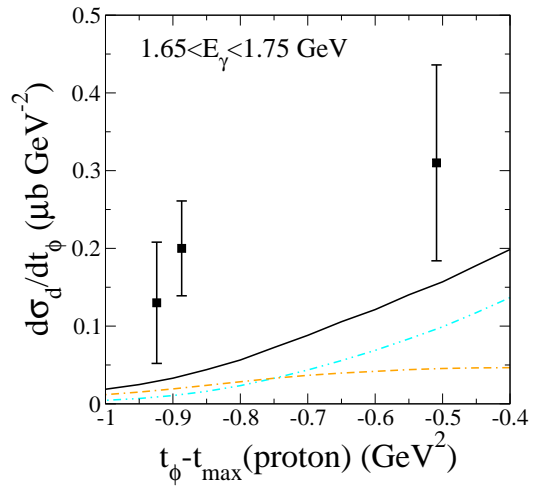}
\end{center}

\caption{Comparison of  our prediction with CLAS data of Ref. \cite{Qian11} for $1.65 < E_\gamma < 1.75$ GeV. Notation for the curves is the same as in Fig. \ref{DCS_1}.}
\label{DCS_CLAS}
\end{figure}

 In Figs. \ref{DCS_1} and \ref{DCS_2}, it is seen that the resonance contribution to the DCS as a function of $t$, given by dash-dash-dotted lines (orange), is basically flat and small. However,
the single-scattering results for DCS with resonance (dash-dotted lines) are significantly larger than those without resonance (dash-dot-dotted lines).
This indicates that the resonant amplitude in general interferes constructively with the nonresonant amplitude. This is in agreement with our findings in
Refs. \cite{Kiswandhi10, Kiswandhi12}. This feature is also seen in Fig. \ref{DCS_CLAS}.
Here we notice in Figs. \ref{DCS_1}, \ref{DCS_2}, and \ref{DCS_CLAS} that the single-scattering results for DCS with resonance (dash-dotted lines)
are significantly larger than those without resonance (dash-dot-dotted lines).
 The effects of the resonance are the most conspicuous at  $|t_\phi - t_{\textrm{max}}$(proton)$| \sim 0.6$ GeV$^2$ where the resonance contribution is about equal to the nonresonant
part and is essential to bring our predictions to agree with the data there which are available only for energy bins of $2.17 < E_\gamma < 2.27$ GeV and $2.27 < E_\gamma < 2.37$ GeV shown in
the lower panel of Fig. \ref{DCS_2}.

\subsubsection{The effects of Fermi motion}
\label{sec:DCS_Fermi}

 In order to understand the role of Fermi motion in the reaction, it is instructive to compare the results of the DCS obtained by
simple summation of free proton and neutron DCS (triangles) and those obtained by using deuteron target without the $pn$ FSI (dash-dotted) in Figs. \ref{DCS_1_medium} and
\ref{DCS_2_medium}. Such a comparison avoids the complication incurred by the presence of  the $pn$ FSI.
It is easily noted that the results are very different at small momentum transfer $|t_\phi - t_{\textrm{max}}$(proton)$|$ and gradually getting more similar at
larger momentum transfer. This feature is  easy to understand.
It is obvious that when the momentum transfer to the deuteron is larger than the momentum of the Fermi motion, the effects from the latter would become less important.
Clearly, at very large momentum transfer, the outgoing struck nucleon would have an absolute velocity large enough such that its initial absolute velocity arising from
Fermi motion would become negligible.

It is also worth mentioning that the results for the DCS of the free nucleons are always above those of the deuteron case without $pn$ FSI.
This is caused by the shape of the energy dependence of the DCS of the $\gamma N \to \phi N$ which decreases sharply toward low energy, but increases slowly toward high energy.
For each value of energy, the Fermi motion will sample this dependence around the energy, and because of the shape of the dependence, the average for the DCS is weighed
toward the low energy, where the change is more drastic, hence its lower values relative to that of the free case. It is also the reason why the differences between the DCS
of the free nucleons and that of the deuteron case are more obvious at lower energy bins.

Unlike the DCS of the free nucleon case which goes to zero drastically at $t_\phi = t_{\textrm{max}}$(proton), the DCS of the $\gamma d \to \phi p n$ reaction goes to
zero more gradually. This is caused also by the Fermi motion of the nucleons inside the deuteron. The nucleon moving at the opposite direction of the photon
provides the reaction the opportunity to produce $\phi$ meson at momentum transfer $t_\phi$ smaller than $t_{\textrm{max}}$(proton) of the free case.

Naively, without considering the internal structure of the deuteron, as well as the $pn$ FSI, one expects that the DCS of $\gamma d \to \phi p n$ reaction is just a sum of the DCS of $\gamma p \to \phi p$ and $\gamma n \to \phi n$. Indeed, when the two mechanisms affect only minimally, for example, at higher energy and larger momentum transfer
$t_\phi$ as one sees from Figs. \ref{DCS_1_medium} and \ref{DCS_2_medium}, the results from $\gamma d \to \phi p n$ are very close to that obtained by summing the DCS of
$\gamma p \to \phi p$ and $\gamma n \to \phi n$.

\begin{figure}[b]

\begin{center}
\includegraphics[width=0.85\linewidth,angle=0]{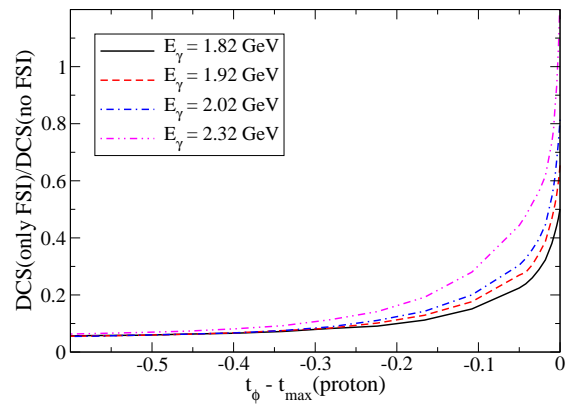}
\end{center}

\caption{The ratio of the DCS with only FSI to the DCS without the FSI at several incoming photon LAB energies.}
\label{DCS_ratio_1}
\end{figure}

\begin{figure}[b]
\begin{center}
\includegraphics[width=1.0\linewidth,angle=0]{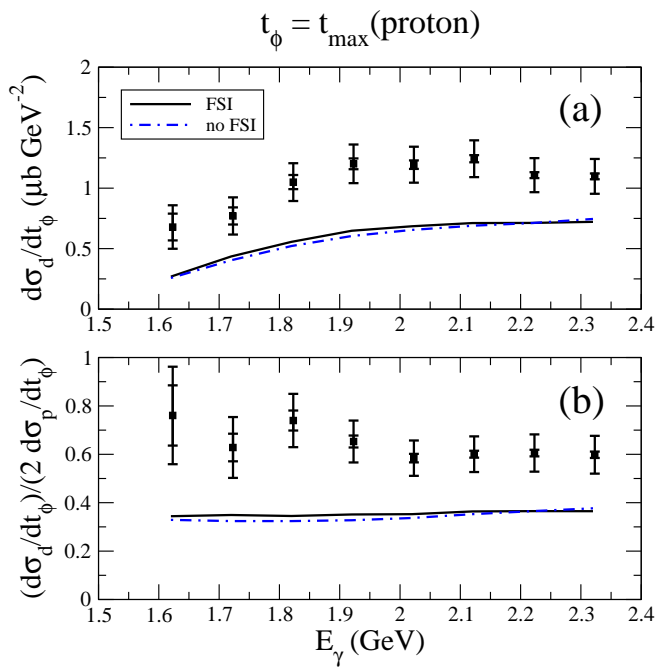}
\end{center}
\vspace{0.0cm}

\caption{(a) The DCS of $\gamma d \to \phi p n$ and (b) the ratio of the DCS of $\gamma d \to \phi p n$ to twice the DCS of $\gamma p \to \phi p$,
both at $t_\phi = t_{\textrm{max}}$(proton) as a function of $E_\gamma$. The notation is as in Fig. \ref{DCS_1} and the dash-dotted lines are the results with resonance but without $pn$
FSI.}
\label{DCS_forward}
\end{figure}

\begin{figure}[b]
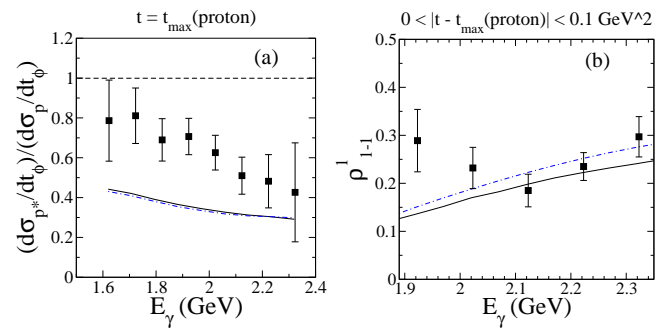

\begin{center}
\includegraphics[width=0.48\linewidth,angle=0]{Figure_10a.eps} \includegraphics[width=0.493\linewidth,angle=0]{Figure_10b.eps}
\end{center}

\vspace{0.0cm}

\caption{(a) The ratio of DCS of $\gamma p \to \phi p$, where the proton is inside the deuteron, to that of free proton, and (b) the SDME $\rho_{1-1}^1$, both as functions of
$E_\gamma$. The notation is as in Fig. \ref{DCS_forward}.}
\label{DCS_forward_1}
\end{figure}

\subsubsection{The effects of FSI}
\label{sec:DCS_FSI}

In Fig. \ref{DCS_1}, it is seen that the differences between results with FSI (full lines) and the one without the inclusion of FSI (dash-dotted lines) are rather small in
the photon energy range of 1.57-1.97 GeV. However, the difference becomes conspicuous in the region $|t_\phi - t_{\textrm{max}}$(proton)$| \le 0.2$ GeV$^2$ as
the photon energy grows larger than 1.97 GeV, as seen in Fig. \ref{DCS_2}. Besides the fact that FSI effects grows with photon energy, a close look further reveals that
FSI effects also increases as $|t_\phi - t_{\textrm{max}}$(proton)$|$ decreases, as depicted in Fig. \ref{DCS_ratio_1}, which gives the ratio of DCS(only FSI)/DCS(no FSI) for the four
energy bins in Figs. \ref{DCS_1} and \ref{DCS_2}.

 The  two distinct features of the FSI effects as mentioned in the above, namely, that they grow  with increasing photon energy and decreasing $|t_\phi - t_{\textrm{max}}$(proton)$|$
are related to the fact that $pn$ cross section drops quickly with increasing energy \cite{PDG2000}.
The reason that FSI effects get magnified with smaller value of $|t_\phi - t_{\textrm{max}}$(proton)$|$ is simply because the invariant mass $M_{pn}$ of $pn$ is monotonically
decreasing with $|t_\phi - t_{\textrm{max}}$(proton)$|$. With a smaller $M_{pn}$, the available CM energy for the $pn$ system, or equivalently
the relative kinetic energy in the CM system of outgoing $pn$ pair also gets smaller such that their interaction becomes stronger.
The growing FSI effects with increasing photon energy can be understood in the same light by noting that the difference between $|t_{\textrm{max}}$(deuteron)$|$
and $|t_{\textrm{max}}$(proton)$|$ becomes smaller in the mean time, as pointed out in Ref. \cite{Sekihara12}.

Another important and interesting feature of the FSI effect is that the inclusion of the final $pn$ rescattering in general brings down the DCS. One first notes  that
the $pn$ FSI actually consist of two parts, the on- and off-shell parts of the integrals over the $pn$ intermediate states as indicated in Eq. (\ref{on-off-shell}).
It can be shown  that the DCS obtained by taking into account only the on-shell part of the $pn$ FSI is actually
equal to the DCS arising from the single-scattering interactions [Fig. \ref{all_diagram}(a)] minus the DCS arising from exclusively the on-shell part of
the $pn$ FSI [Fig. \ref{all_diagram}(b)]. This interesting result can be understood  theoretically as
a consequence of the unitarity of the $pn \to pn$ amplitude and is independent of the details of the interaction. The proof is given in Appendix \ref{app:DCS_relation}.
At $E_\gamma \sim 2$ GeV and $|t_\phi - t_{\textrm{max}}$(proton)$| < 0.05$ GeV$^2$, the DCS with on-shell $pn$ FSI (dashed), shown in Figs. \ref{DCS_1_medium}
and \ref{DCS_2_medium}, are around $40 \%$ lower than the results without the $pn$ FSI. The large on-shell $pn$ FSI effects contradicts the results of  Ref. \cite{Sekihara12},
where it is found to be small. This can be understood as a consequence of the fact that their $\gamma d \to \phi pn$ and $pn \to pn$ amplitudes are not unitary,
This  suggests that unitarity of the reactions taking place in the final state should be taken into account in a study that includes FSI.

The inclusion of the off-shell FSI brings up the DCS again as seen in Figs. \ref{DCS_1_medium} and \ref{DCS_2_medium}, namely, the effects of on-shell and off-shell FSI
cancel out each other to some extent.
In the end, the total $pn$ FSI results (full)  in general lie between the results with only on-shell FSI included (red dashed lines) and those without $pn$ FSI (blue
dash-dotted lines). For example, at $E_\gamma \sim 2$ GeV and $|t_\phi - t_{\textrm{max}}$(proton)$| < 0.05$ GeV$^2$, the final DCS are only
around $20 \%$ lower than the results without $pn$ FSI, after both the on-shell and off-shell FSI are considered. This demonstrates that the off-shell FSI are
important as they actually reduce the effects introduced by the on-shell ones up to about  50$\%$  at these kinematics.

\subsubsection{Comparison with the data}
\label{sec:DCS_data}

 We next compare  our results for the DCS with the LEPS data \cite{Chang10DCS, Chang_pc}. Here, in Figs. \ref{DCS_1} and \ref{DCS_2}, our results without both
resonance and $pn$ FSI, as given by the dash-dot-dotted line, show a strong peaking tendency at small values of $|t_\phi - t_{\textrm{max}}$(proton)$|$, especially at higher energies,
as indicated by the data. However, they are in general smaller than the data, except at the peak region of  $|t_\phi - t_{\textrm{max}}$(proton)$| < 0.05$ GeV$^2$ at the highest energy bin of
$2.27 < E_\gamma < 2.37$ GeV considered in this study.
After the  inclusion of the resonance contribution, but not $pn$ FSI effects, the results (dash-dotted line) are all shifted by roughly the same amount for all $|t_\phi - t_{\textrm{max}}$(proton)$|$.
The inclusion of the resonance  improves the agreement with  data at lower energies but overshoots the data at
small values of $|t_\phi - t_{\textrm{max}}$(proton)$|< 0.05$ GeV$^2$ at higher energies. Lastly, when the $pn$ FSI is included in addition to the resonance, the agreement with the data
improves as the FSI significantly reduces the peak.

 Nevertheless, some discrepancies remain at small momentum transfer, namely, in the peak region, especially at photon  energies higher than 2.17 GeV,
where at the highest energy bin, a difference of about $25\%$ is observed.
On the other hand, our prediction for the peak in the lowest energy bin $1.57 < E_\gamma < 1.67$ GeV is
considerably lower than the data. In the end, with all the effects included,
our DCS results underestimate the data at lower energies while overestimate them at higher energies. Notice also that our DCS results explain the LEPS data well at larger
momentum transfer but not at small momentum transfer.
However, when comparing  with the CLAS data of Ref. \cite{Qian11}, which is taken at large momentum transfer
of $-0.9 <t_\phi - t_{\textrm{max}}$(proton)$< -0.5$ GeV$^2$ in the   energy bin of $1.65 < E_\gamma < 1.75$ GeV,   as shown in Fig. \ref{DCS_CLAS}, we find that our predictions are lower than the data.

 It is   to be noted that at $t_\phi$ very close to $t_{\textrm{max}}$(proton), the LEPS data do not
fall to zero while our results do. It is suggested by the main author \cite{Chang_pc} of Ref. \cite{Chang10DCS} that it is possible that,
due to the binning size of the data, the sharp decrease of the DCS around $t_\phi = t_{\textrm{max}}$(proton) might not have been represented well in the experimental results.

 The comparison of our results for the $d\sigma_d/dt_\phi$ and $(d\sigma_d/dt_\phi)/(2 d\sigma_p/dt_\phi)$ at forward direction $t_\phi = t_{\textrm{max}}$(proton) to the data
of Ref. \cite{Chang10DCS} is presented in Fig. \ref{DCS_forward}. In Fig. \ref{DCS_forward_1}(a), we also compare our results of the ratio of
$(d\sigma_{p^*}/dt_\phi)/(d\sigma_p/dt_\phi)$ to the data of Ref. \cite{Chang10DCS}, where $d\sigma_{p^*}/dt_\phi$ is the DCS of $\gamma p \to \phi p$ in which the
$\phi$ meson is produced from the proton inside the deuteron. Here, we observe that our results do not match the data, even though the shape are generally in agreement with the
data. As these results are taken at forward direction, it is possible that the problem is related to the size of the bins as discussed above. Lastly, it is important to point
out here that in both the data and our results, no clear peak corresponding to the resonance shows up in Fig. \ref{DCS_forward}(a). This is readily understandable as the resonance
is weak and easily gets smeared out by the Fermi motion and FSI with the deuterium target.

\subsection{Spin-density matrix elements}
\label{sec:SDME}

\begin{figure}[h]
\begin{center}
\includegraphics[width=1.0\linewidth,angle=0]{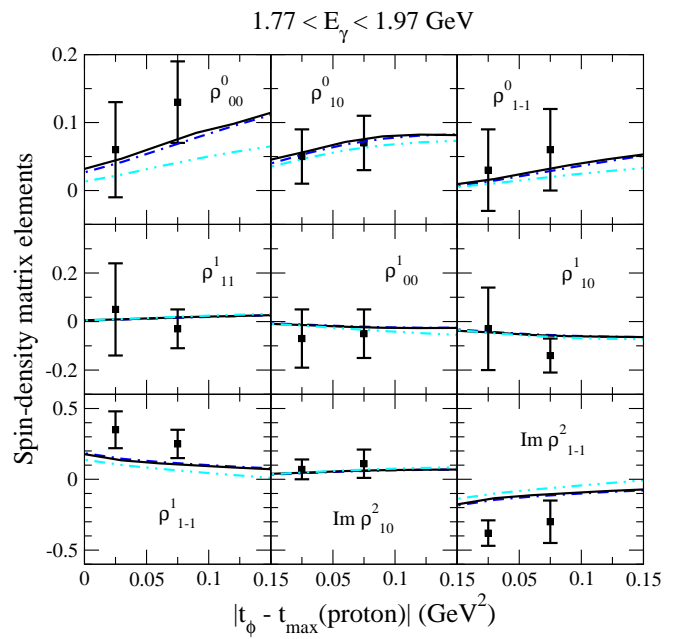}
\end{center}

\vspace{-0.5cm}

\caption{The results for the SDME of $\gamma d \to \phi p n$ reaction for $1.77 < E_\gamma < 1.97$ GeV with resonance and $pn$ FSI, with resonance but without $pn$
FSI, and without both resonance and $pn$ FSI are given by full, dash-dotted, and dash-dot-dotted lines, respectively. The data are from Ref. \cite{Chang10}.}
\label{SDME_1}
\end{figure}

\begin{figure}[h]
\begin{center}
\includegraphics[width=1.0\linewidth,angle=0]{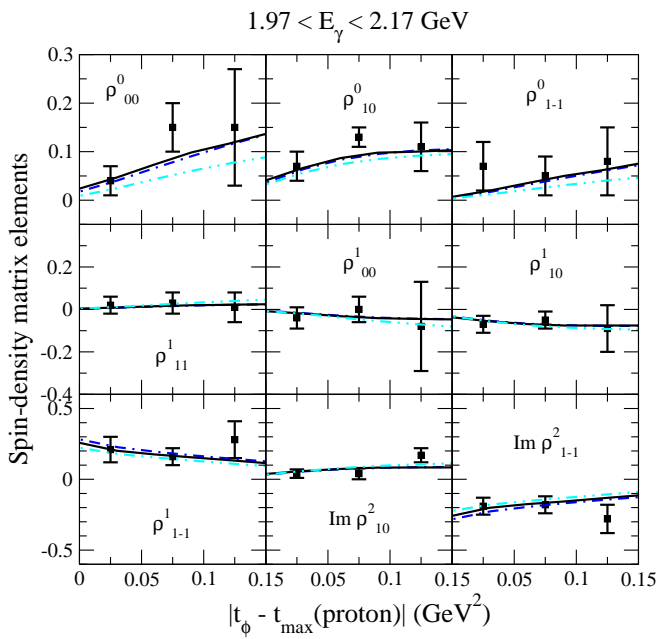}
\end{center}

\vspace{-0.5cm}

\caption{Caption is the same as in Fig. \ref{SDME_1}, but for $1.97 < E_\gamma < 2.17$ GeV.}
\label{SDME_2}
\end{figure}

\begin{figure}[h]
\begin{center}
\includegraphics[width=1.0\linewidth,angle=0]{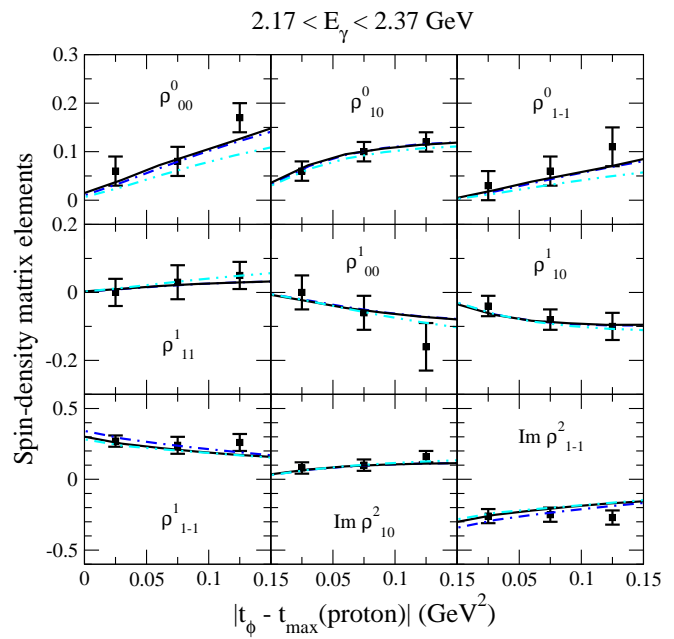}
\end{center}
\caption{Caption is the same as in Fig. \ref{SDME_1}, but for $2.17 < E_\gamma < 2.37$ GeV.}
\label{SDME_3}
\end{figure}

 The   SDME as a function of $t_\phi$ from LEPS at three energy bins \cite{Chang10} are shown in Figs. \ref{SDME_1}, \ref{SDME_2}, and \ref{SDME_3}, and the SDME
$\rho_{1-1}^1$ as a function of photon LAB energy are presented in Fig. \ref{DCS_forward_1}(b). The notation is the same as in Figs. \ref{DCS_1}-\ref{DCS_1_medium}. Namely, full, dashed-dotted, and dash-dot-dotted lines
denote results obtained with resonance and FSI included, only resonance but no FSI included, and without both resonance and FSI included, respectively. We will first elaborate on the role
of the spin dependence of the elementary amplitude, as well as the roles of resonance, and FSI, before discussing the comparison with data.

\subsubsection{The spin dependence of the elementary amplitude of $\gamma N\rightarrow \phi N$}
\label{sec:SDME_spin}

 The proper spin dependence of the elementary $\gamma N \to \phi N$ amplitude is employed in our calculation whereas its dependence
on the nucleon spin was simply neglected  in Ref. \cite{Sekihara12}.
  Without the use of such spin dependence, it is not possible to describe the SDME data as all of them would be zero except for
$\rho_{1-1}^1 = -\textrm{Im} \rho_{1-1}^2 = 0.5$.

\subsubsection{The role of the resonance}
\label{sec:SDME_res}

 For the SDME,  we observe in Figs. \ref{SDME_1}, \ref{SDME_2},  and \ref{SDME_3}, that the inclusion of the resonance does not affect as much as in the case of DCS,
as seen in the difference between dashed-dotted, and dash-dot-dotted lines, except the $\rho_{00}^0$. This means that it does not
change much the polarization properties of the reaction. This is a consequence of the normalization of the SDME being proportional to the DCS.
However, in terms of percentage,  the  $\rho_{1-1}^0$, $\rho_{11}^1$, $\rho_{1-1}^1$, and $\textrm{Im} \rho_{1-1}^2$ elements are all rather
significantly changed, especially at larger momentum transfer, in agreement to what we found in Ref. \cite{Kiswandhi12}.

 The enhancement of the $\rho_{00}^0$ element results from the increase on the production of the $\phi$ meson of helicity $\lambda_\phi = 0$.
On the other hand, the enhancement of the $\rho_{11}^1$ and $\rho_{1-1}^1$ elements indicates that double-spin-flip transition with $\lambda_\phi = -\lambda_\gamma$
increases.
This is actually a consequence of the fact that the resonance process is not helicity-conserving, in contrast to the Pomeron-exchange process which is
basically helicity-conserving. The enhancement of the $\rho_{1-1}^1$ and $\textrm{Im} \rho_{1-1}^2$ elements basically shows that the resonance does increase the
natural-parity production which is already strongly reduced by the $\pi$- and $\eta$-exchange processes.

It is important to note that, even for the elements where the resonance does not seem to affect much, it is misleading
to think that it does not contribute. As explained in Ref. \cite{Kiswandhi12}, the resonance does contribute, but however, the interference between
resonance and nonresonance processes somewhat balances the resonance contribution in the opposite direction.

\subsubsection{The effects of FSI}
\label{sec:SDME_FSI}

 In Figs. \ref{SDME_1}, \ref{SDME_2},  and \ref{SDME_3}, it is observed that the full and dash-dotted lines almost
sit upon each other which implies that the FSI effects on the SDME are minimal, even at the highest energy bin
$2.17 < E_\gamma < 2.37$ GeV, where  the FSI effect on the DCS has been substantial.
This is a consequence of the fact that the SDME are actually normalized by the amplitude square. It also reflects
the fact that the final $pn$ rescattering does not change the spin distribution of the produced $\phi$ meson.

 The only two SDME significantly changed by the $pn$ FSI are $\rho_{1-1}^1$ and $\textrm{Im} \rho_{1-1}^2$
at small $|t_\phi - t_{\textrm{max}}$(proton)$|$ values. The reductions from this effect are around $15 \%$ in the highest energy bin. This basically shows that the $pn$ FSI actually increases the strength of the unnatural-parity exchange mediated by the intermediate $\pi$ and $\eta$ relative to that of
natural-parity exchange by the pomeron. This can be understood readily. First, consider the case without the $pn$ FSI.
Here, notice that the strength of the unnatural-parity exchange in the incoherent case would be lower than the free proton one since in the former, the final
proton and neutron moving at similar velocities would cause their $\pi$-exchange amplitudes to interfere destructively.
However, the $pn$ FSI decreases the production of the final proton and neutron moving at similar velocities, which enhances the strength of unnatural parity exchange and
manifested in the smaller values of the two SDME mentioned in the above.

Finally, it is also interesting to note that the reduction on $\rho_{1-1}^1$ and $\textrm{Im} \rho_{1-1}^2$ due to rescattering of final $pn$ is not caused by stronger
double-spin-flip transition where $\lambda_\phi = -\lambda_\gamma$ in which $\lambda_\gamma$ ($\lambda_\phi$) is the helicity of photon ($\phi$ meson) although these two SDME
contain contributions from this transition. It can be seen from Figs. \ref{SDME_1}, \ref{SDME_2}, and \ref{SDME_3} in which the SDME $\rho^0_{1-1}$ and $\rho^1_{11}$,
which are  measures of the strength of the double-spin-flip transition,  basically almost vanish  at small $|t_\phi - t_{\textrm{max}}$(proton)$|$ values.

\subsubsection{Comparison with the data}
\label{sec:SDME_data}

We observe that the presence of the resonance helps improves the agreement with data for $\rho_{00}^0$ at all energies.
However, the resonance does not help the description of $\rho_{1-1}^1$ and $\textrm{Im} \rho_{1-1}^2$ at the lowest energy bin $1.77 < E_\gamma < 1.97$ GeV.
As for the inclusion of the $pn$ FSI, we observe that it does not bring any significant betterment to the agreement of our results with the data.
After including all the effects, we notice that the description of $\rho_{1-1}^1$ and $\textrm{Im} \rho_{1-1}^2$ at the lowest energy bin $1.77 < E_\gamma < 1.97$ GeV
is still off from the data. Similarly,  our results for the energy dependence of $\rho_{1-1}^1$ at small momentum transfer of $|t_\phi - t_{\textrm{max}}$(proton)$|< 0.1$ GeV$^2$
in Fig. \ref{DCS_forward_1}(b)  describe the data well, but poor at lower energies.
This indicates  that the data favors more contribution from natural parity exchange which is usually provided by Pomeron exchange.

\subsection{Discussions}
\label{sec:discussions}

 From the comparisons between our predictions with the data as presented in Secs. \ref{sec:DCS_data} and \ref{sec:SDME_data}, it is seen that the overall agreement is satisfactory, namely,
the main features of the data are all properly reproduced except for a few quantitative discrepancies.

 Regarding the DCS, the most serious
discrepancies between the data and our results lie in the neighborhood of the forward direction, i.e., small values of $|t_\phi - t_{\textrm{max}}$(proton)$|$,
as seen in Figs. \ref{DCS_1} and \ref{DCS_2}. One notices that  the height of the predicted peak is at first underpredicted at the lower energy bins with the difference getting smaller with
increasing energy and then eventually overshoots the data. Yet at the very forward direction with $t_\phi = t_{\textrm{max}}$(proton), our predictions
are always smaller than the data as seen in Fig. \ref{DCS_forward}(a).  It might be related to the binning size of the data \cite{Chang_pc} as discussed in Sec. \ref{sec:DCS_data}. For the intermediate momentum transfer region with $0.2<|t_\phi - t_{\textrm{max}}$(proton)$|< 0.5$ GeV$^2$, our results
describe well the data. But as momentum transfer grows and becomes larger than $0.5<|t_\phi - t_{\textrm{max}}$(proton)$|$ GeV$^2$, our results underestimate the CLAS data of \cite{Qian11}
as indicated in Fig.  \ref{DCS_CLAS}. However, it should be noted that this data is taken at a rather  low  energy bin of $1.65 < E_\gamma < 1.75$ GeV among the data set we consider in this study.

 The SDME measured by LEPS are in general reproduced well by our calculations besides $\rho_{1-1}^1$ and $\textrm{Im} \rho_{1-1}^2$
in the low energy bin of $1.77 < E_\gamma < 1.97$ GeV. This happens to be the case in our model for the elementary amplitude of $\gamma p\rightarrow \phi p$ \cite{Kiswandhi12}.

 The discrepancies summarized above are characterized  by (i) very low energies for all momentum transfers, and (ii) forward angles except for the energy bins lying between
1.87 GeV and 2.17 GeV.

 It is then an interesting question to ask why our model, which has performed well in the $\gamma p \to \phi p$ case does not describe
 $\gamma d \to \phi p n$ as well as one would expect despite we have treated the problem laboriously up to the double scattering contributions.
 The higher order multiple rescatterings as depicted in Figs. \ref{all_diagram}(d) and (e) are likely not to be blamed since they all involve at least one
 meson-nucleon scattering and hence would not contributed significantly as discussed in Sec. \ref{sec:gamma_d_double}. The only possibility left is then
 the model we employ for the elementary process $\gamma N\rightarrow \phi N$. This is also supported by noting that
   the disagreement found in the SDME is somewhat  similar to that in the
free proton case presented in Ref. \cite{Kiswandhi12}.

Three possible causes have come to our mind. Namely, (i) description of the resonance, (ii) off-shell behavior of the elementary amplitude, and (iii) low energy behavior
of the model.

Discussing the role of the resonance in  $\gamma d \to \phi p n$ reaction,  one first notes that in order to study the $\gamma d \to \phi p n$ reaction, a model for the $\gamma n \to \phi n$ reaction is required.
The background part of the Pomeron, $\pi$, and $\eta$ exchanges can be readily obtained with isospin invariance. Regarding the resonance
contribution, there is a problem in that there exists no data for
$\phi$ production from neutron  to infer the couplings of $\gamma nN^*$ and $\phi nN^*$ such that
 we have to rely on some model assumptions for estimation.  However, we find from Figs. \ref{DCS_1} and \ref{DCS_2}  that
  the resonance contribution is   rather independent of momentum transfer as in the case of $\gamma p \to \phi p$ \cite{Kiswandhi12}, while
  discrepancies appear to be the largest at small momentum transfer. Accordingly, an improved description of the resonance
  may not be sufficient to bridge the discrepancy.   Nevertheless, it has to be reminded that resonance does play a significant role
in reducing the difference with the data, especially for the DCS and some elements of the SDME at larger momentum transfer. Another interesting
question is whether some peak feature would appear in $\gamma d \to \phi p n$, as in $\gamma p \to \phi p$. In this connection, it is to be reminded that,
according to our analysis in Ref. \cite{Kiswandhi12}, the nonmonotonic behavior found in Ref. \cite{leps05} is a result of a subtle balance between the meson exchange mechanism
and a weak resonance. Such a balance could easily be offset by the Fermi motion and FSI in the $\phi$ production from deuteron.

In Figs. \ref{DCS_1_medium} and \ref{DCS_2_medium}, it can be observed that our DCS results with resonance and on-shell rescattering only (dashed lines) fit the data very well
at higher energies, though  not as well at lower energies. This may be related to the possibility that the off-shell rescattering contribution to the DCS is not satisfactorily
 estimated. This is indeed an open question as the pomeron amplitude itself, which constitutes the largest contribution to the DCS, is not yet fully understood, not mentioning
its off-energy-shell behavior as is needed in our calculation. This question should be investigated in more details.

The low energy behavior of our model, basically that of the Pomeron amplitude, as well as contributions from other   mechanisms like existence of second Pomeron, glueball exchange, and $s\bar s$ knockout etc., has been the subject of several studies, e.g.,
Refs. \cite{titov03,titov07a}. Again they remain to be studied in depth.

 The issues enumerated above could possibly be answered by the recent high statistics data on the $\gamma p \rightarrow \phi p$ DCS and SDME taken at CLAS \cite{Dey14,Sera14}. Their DCS data, taken at a larger range of momentum transfer and energy, also confirm  the nonmonotonic behavior observed at small momentum transfer previously by LEPS. However, their data indicate that the nonmonotonic behavior
does not appear at larger momentum transfer. This, in fact, could possibly cast some doubts on
whether the nonmonotonic behavior is really caused by the presence of a resonance.    We have recently realized that $s \bar{s}$ knockout mechanism as considered in Ref.
\cite{titov97a} could also produce nonmonotonic behavior in the forward differential cross section of $\gamma p \rightarrow \phi p$ and  are currently engaged in an attempt to extend
our model  for  $\gamma p \to \phi p$  \cite{Kiswandhi12} by including the $s \bar{s}$ knockout process to see whether it is possible to explain the LEPS data \cite{leps05,Chang10} and the recent CLAS data \cite{Dey14,Sera14}. It will conceivably shed useful light on the issues discussed above.

\vskip 0.5cm

\section{Summary and Conclusions}
\label{sec:conclusions}


In summary, we have calculated the differential cross sections (DCS) and spin-density matrix elements (SDME) of the incoherent photoproduction of $\phi$ meson from deuteron $\gamma d \to \phi p n$ near threshold and
compared them with the  data from the LEPS \cite{Chang10,Chang10DCS} and CLAS \cite{Qian11}. The calculation is based on a model for
$\gamma p \to \phi p$ we constructed in Refs. \cite{Kiswandhi10, Kiswandhi12} which described the LEPS data of Ref. \cite{leps05} well,  including
a peak around photon lab energy of 2.0 GeV, first observed by the LEPS Collaboration, and  recently confirmed by the high statistics CLAS data \cite{Dey14,Sera14}. Our
model contains of a resonance with spin-parity $J^P = 3/2^-$, mass $M_{N^*}=2.08\pm 0.04$ GeV, and width $\Gamma_{N^*} =0.570\pm 0.159$,
in addition to a background consisting of Pomeron and $(\pi, \eta)$ exchanges in the $t-$channel. For production from deuteron, the couplings of the resonance to neutron are
estimated with a guide from relativistic constituent quark model.
The calculation for $\gamma d \to \phi p n$ was then carried out up to double rescatterings with realistic nucleon-nucleon interaction.

Our calculation contains  major improvements over  that of Ref. \cite{Sekihara12} in that (i) the Pomeron
amplitude used is more realistic and contains the proper spin dependence,  which leads to nontrivial values for the SDME; (ii) the $D$ state of deuteron is considered, and (iii) the off-shell
contributions to the FSI of the final $pn$, in addition to the on-shell part, are included. Accordingly, our results are, up to double rescatterings, predictions
of  $\gamma d \to \phi pn$ reaction as no more fitting is involved.

We find that the inclusion of Fermi motion of the nucleons inside the deuteron plays an important role in the description of one characteristic of   $\gamma d \to \phi p n$
at small momentum transfer, namely, a gradual decrease of DCS with diminishing $|t_\phi - t_{\textrm{max}}$(proton)$|$. On the other hand, the contribution from deuteron
$D$ state to both DCS and SDME is negligible.

The effects of FSI involving meson-nucleon rescatterings is estimated to be small. However, the effects from the rescattering of the final $pn$ are found to be rather significant to provide a considerable   reduction in
the DCS which is about $20 \%$ at higher energy bins. We find that the contributions of on-shell and off-shell final $pn$ rescatterings
cancel out each other to some extent. The off-shell rescattering effects should hence be considered in realistic description of the reaction. It brings up the question on the off-shell extrapolation of the Pomeron amplitude, an issue remains to be studied further.

Regarding the postulated resonance, it is found to play a significant role in the description of the DCS at a broad range of
momentum transfer and energy,  although less so for the SDME. The weak resonance, which is responsible for the appearance of a small peak in $\gamma p\rightarrow\phi p$,
is not found to produce any nonmonotonic behavior in $\gamma d \rightarrow \phi p n$, apparently smeared out by the Fermi motion and FSI.

The overall agreement of our results with the data is satisfactory in that the main features of the data are all properly reproduced except a few quantitative discrepancies
at some kinematic regions characterized by (i) very low energies at all momentum transfers, and (ii) forward angles except at the energy bins lying between
1.87 GeV and 2.17 GeV.

On the experimental side, the DCS data from LEPS do not fall to zero at small $t$ as would be expected. According to Ref. \cite{Chang_pc}, it might be related
to the binning size of the data. It is possible that some aspects of their results might need to be further examined.
In addition, the LEPS data have relatively large error bars as they had to rely on MC simulation to separate the coherent and incoherent events.
The comparison of our predictions with the data will be more meaningful after these questions are clarified.

On the theoretical side, our model for the elementary amplitude of $\gamma p \to \phi p$ should be carefully compared with the recent high
statistics data from CLAS \cite{Dey14,Sera14}, especially regarding the assumption of resonance.  Other possible mechanisms like
nondiffractive processes of  nucleon exchange, nucleon
resonances, second Pomeron exchange,
$t$-channel scalar meson and glueball exchanges, and $s\bar s$-cluster knockout
  which have been suggested to contribute at low energies, should also
be re-examined, in light of the new CLAS data.     We  recently realize that $s \bar{s}$ knockout mechanism as studied in Ref.
\cite{titov97a} could also produce nonmonotonic behavior in the forward differential cross section of $\gamma p \rightarrow \phi p$ and  are currently engaged in an attempt to extend
our model  for  $\gamma p \to \phi p$  \cite{Kiswandhi12} by including the $s \bar{s}$ knockout process to see whether it is possible to account for the LEPS data \cite{leps05,Chang10} and the recent high statistics CLAS data \cite{Dey14,Sera14}.
  Finally, we like to emphasize
that a combined analysis of the low energy data of $\gamma p \to \phi p$ and
 $\gamma d \to \phi p n$ will be much desired to shed light on the issues discussed in the above.

\acknowledgments
We would like to thank Dr. Wen Chen Chang for useful discussions and correspondences.
This work was supported in parts by National Science Council of the
Republic of China (Taiwan) under grant NSC100-2112-M002-012. We
would also like to acknowledge the help from National Taiwan
University High-Performance Computing Center in providing us with a
fast and dependable computation environment which is essential in
carrying out this work. A. K. would like to thank IHEP, Beijing, Center of Theoretical Science, National
Taiwan University,  and National Center for Theoretical Sciences,
  Hsin-Chu, Taiwan, for supports of visits to the respective institution
during the course of this study. Y. B. D. also thanks the Sino-German CRC 110 ``Symmetries and the Emergence of Structure in QCD" project (NSFC Grant No. 11261130311), and the
Center of Theoretical Sciences, National Taiwan University for their hospitalities.

\appendix

\section{The model for $\gamma N \to \phi N$ reaction}
\label{app:model}
We first introduce the kinematic variables $k$, $p_i$, $q$, and
$p_f$ for the four-momenta of the incoming photon, initial proton,
outgoing $\phi$-meson, and final proton, respectively, with
$s=(k+p_i)^2=(q+p_f)^2$, $t=(q-k)^2 = (p_f-p_i)^2$, and
$u=(p_f-k)^2=(q-p_i)^2$.

\begin{figure*}
\includegraphics[width = 0.8\linewidth,angle=0]{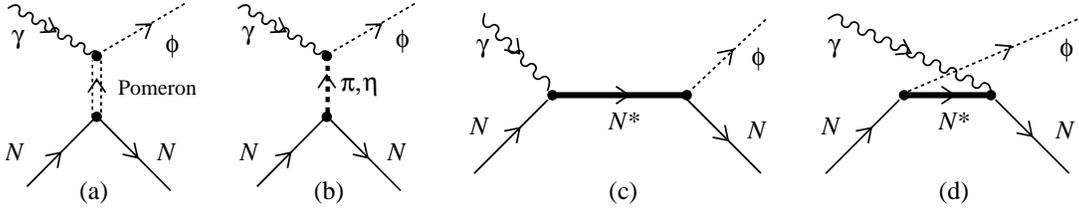}
\caption{Pomeron, $(\pi,\eta)$ exchanges, $s$-, $u$-channel $N^*$ excitation diagrams for $\gamma N \to \phi N$ reaction are labeled (a), (b), (c), and (d), respectively.}
\label{gammapdiagram}
\end{figure*}

In addition to the nonresonant mechanism of Pomeron-exchange, $t$-channel $\pi$-
and $\eta$-exchange, we include a resonance $N^*$. We can then write the full amplitude ${\cal
M}$ as
\begin{eqnarray}
{\cal M}_{\gamma N \to \phi N} ={\cal M}_P + {\cal M}_{\pi + \eta} + {\cal M}_{N^*},
\label{amp}
\end{eqnarray}
as shown in Fig.~\ref{gammapdiagram}, where ${\cal M}_{N^*}$
contains both $s$- and $u$-channel contributions.

\subsection{Pomeron exchange}
\label{sec:pomeron}
Following Refs.~\cite{titov03,titov07}, we can easily write down the
Pomeron-exchange amplitude of Fig.~\ref{gammapdiagram}(a) ,
\begin{eqnarray}
{\cal M}_P &=& -\bar{u}(p_f,\lambda_{N'})M(s,t)\Gamma^{\mu \nu} u(p_i,\lambda_N) \nonumber \\
&\times&\varepsilon^*_{\mu}(q,\lambda_{\phi})\varepsilon_{\nu}(k,\lambda_{\gamma}),
\label{ampa}
\end{eqnarray}
where $\varepsilon_{\mu}(q,\lambda_{\phi})$ and
$\varepsilon_{\nu}(k,\lambda_{\gamma})$ are the polarization vectors
of the $\phi$-meson and photon with  $\lambda_{\phi}$ and
$\lambda_{\gamma}$, respectively, and
$u(p_i,\lambda_N)$[$u(p_f,\lambda_{N'})$] is the Dirac spinor of the
nucleon with momentum $p_i$($p_f$) and helicity
$\lambda_N$($\lambda_{N'}$). The transition operator $\Gamma^{\mu
\nu}$ in Eq. (\ref{ampa}) is
\begin{eqnarray}
\Gamma^{\mu \nu} &=& \left(g^{\mu \nu}-\frac{q^{\mu} q^{\nu}}{q^2}\right)
\not\!k - \left(k^{\mu} - \frac{k \cdot q q^{\mu}}{q^2}\right)
\gamma^{\nu} \nonumber \\
&-& \left(\gamma^{\mu}-\frac{\not\!q
q^{\mu}}{q^2}\right)\left[q^{\nu} - \frac{k \cdot
q(p^{\nu}_i+p^{\nu}_f)}{k \cdot (p_i + p_f)}\right].
\end{eqnarray}
The scalar function $M(s,t)$ is described by the Reggeon parametrization,
\begin{eqnarray}
M(s,t) &=& C_P
F_1(t)F_2(t)\frac{1}{s}\left(\frac{s-s_{th}}{s_0}\right)^{\alpha_P(t)} \nonumber \\
&\times&\text{exp}\left[-\frac{i\pi}{2}\alpha_P(t)\right],
\end{eqnarray}
where we have taken the Pomeron trajectory $\alpha_P(t) = 1.08 +
0.25 t$ and $s_0=(M_N+M_\phi)^2$.   The isoscalar form
factor $F_1(t)$ of the nucleon and the form factor $F_2(t)$ of the
$\phi$-photon-Pomeron coupling are given as
\cite{titov03,donnachie87},
\begin{eqnarray}
F_1(t)&=& \frac{4M^2_N-a_N^2t}{(4M^2_N-t)(1-t/t_0)^2},\\
F_2(t)&=& \frac{2\mu^2_0}{(1-t/M^2_{\phi})(2\mu^2_0+M^2_{\phi}-t)},
\end{eqnarray}
with $\mu^2_0 = 1.1$ $\textrm{GeV}^2$, $a_N^2 = 2.8$, and $t_0 = 0.7$ $\textrm{GeV}^2$.

Here, the strength factor is taken to be $C_P = 3.65$ which is obtained by fitting to the
total cross sections data at high energy \cite{titov03}. Included as well is the threshold factor $s_{th}$ \cite{williams98,titov03} so that we get a better agreement
with  experimental data near the threshold region. Due to pomeron properties and behaviors at lower energies are
not well-established, we adjust this parameter to fit the experimental data on the DCSs around
$E_\gamma = 6$ GeV because at this energy, it can be expected that all other contributions from
hadronic intermediate states would become negligible and only pomeron contributes. Also, around this energy, experimental
data have relatively small error bars and rise steadily without much fluctuation. These give us
confidence to match the pomeron contribution to the experimental data at this energy by fixing $s_{th} = 1.3$ ${\textrm{GeV}^2}$.

\subsection{$\pi$ and $\eta$-meson exchanges}

The amplitudes for the $\pi$ and $\eta$ exchanges in $t$ channel,
Fig.~\ref{gammapdiagram}(b) can be calculated straightforwardly
\cite{titov97,titov98} and are given by
\begin{eqnarray}
{\cal M}_{\pi + \eta} &=& \frac{-eg_{\gamma\phi \pi}g_{\pi
NN}F^2_{\pi}(t)}{M_{\phi}}\bar{u}(p_f,\lambda_{N'})\gamma_5\frac{\varepsilon^{\mu\nu\rho\sigma}q_{\mu}k_{\rho}}{t-M^2_{\pi}} \nonumber \\
&\times& u(p_i,\lambda_N)\varepsilon^*_{\nu}(q,\lambda_{\phi})\varepsilon_{\sigma}(k,\lambda_{\gamma})+
\nonumber \\
&-& \frac{eg_{\gamma\phi \eta}g_{\eta
NN}F^2_{\eta}(t)}{M_{\phi}}\bar{u}(p_f,\lambda_{N'})\gamma_5\frac{\varepsilon^{\mu\nu\rho\sigma}q_{\mu}k_{\rho}}{t-M^2_{\eta}} \nonumber \\
&\times& u(p_i,\lambda_N)\varepsilon^*_{\nu}(q,\lambda_{\phi})\varepsilon_{\sigma}(k,\lambda_{\gamma}),
\label{ampb}
\end{eqnarray}
with the coupling constants $g_{\pi NN} = 13.26$, $g_{\gamma \phi \pi} = -0.14$, and
$g_{\gamma \phi \eta} = -0.71$, as well as the form factors $F_{\pi}(t)$ and
$F_{\eta}(t)$ for the virtually exchanged mesons at the $MNN$ and
$\gamma \phi M$ ($M=\pi,\eta$) vertices, respectively, are taken to
be the same as in Ref.~\cite{titov07}. We choose $g_{\eta NN} =
1.12$~\cite{wentai} and $\Lambda_{\pi} = \Lambda_{\eta} = 1.2$ GeV
which are slightly different with the values given in
Ref.~\cite{titov07}.

\subsection{Excitation of a baryon resonance}

The $s$- and $u$-channel Feynman diagrams with an $N^*$ in the intermediate state
are shown in Fig.~\ref{gammapdiagram}(c) and (d).
For the coupling of $3/2$ resonances to $\gamma N$, we choose the commonly used
interaction Lagrangians~\cite{wentai,feu,pascalutsa}
\begin{eqnarray}
{\cal L}_{\gamma N N^*}^{3/2^\pm} &=& i e g^{(1)}_{\gamma N N^*} \bar{\psi}_{N} \Gamma^\pm \left(\partial^\mu\psi_{N^*}^\nu\right) \tilde{F}_{\mu \nu} \nonumber \\
&+& eg^{(2)}_{\gamma N N^*} \bar{\psi}_{N} \Gamma^\pm \gamma^5 \left(\partial^{\mu} \psi_{N^*}^\nu \right)F_{\mu \nu} + \textrm{h.c.}, \label{photonthreehalf}
\end{eqnarray}
where $F_{\mu \nu}= \partial_{\mu}A_{\nu} - \partial_{\nu}A_{\mu}$
is the electromagnetic field tensor, and $\sigma_{\mu \nu} =
\frac{i}{2}(\gamma_{\mu}\gamma_{\nu}-\gamma_{\nu} \gamma_{\mu})$.
Also, $\tilde{F}_{\mu \nu} = {1 \over 2}
\epsilon_{\mu\nu\alpha\beta} F^{\alpha\beta}$ denotes the dual
electromagnetic field tensor with $\epsilon^{0123} = +1$. The
operator $\Gamma^\pm$ are given by $\Gamma^+=1$ and
$\Gamma^-=\gamma_5$. For the $\phi N N^*$ interaction Lagrangians,
we have
\begin{eqnarray}
{\cal L}_{\phi N N^*}^{3/2^\pm} &=& ig^{(1)}_{\phi N N^*}
\bar{\psi}_{N} \Gamma^\pm \left(\partial^\mu\psi_{N^*}^\nu\right) \tilde{G}_{\mu \nu} \nonumber \\
&+& g^{(2)}_{\phi N N^*} \bar{\psi}_{N} \Gamma^\pm \gamma^5
\left(\partial^{\mu} \psi_{N^*}^\nu \right)G_{\mu \nu} \nonumber \\
&+& ig^{(3)}_{\phi N N^*} \bar{\psi}_{N} \Gamma^\pm \gamma^5  \gamma_\alpha
\left(\partial^\alpha \psi^\nu_{N^*} - \partial^\nu \psi^\alpha_{N^*}\right)
\left(\partial^\mu G_{\mu\nu}\right) \nonumber \\
&+& \textrm{h.c.}, \label{phiNNstar}
\end{eqnarray}
where   $G^{\mu \nu}$ is defined as $G^{\mu \nu} =
\partial^{\mu} \phi^{\nu}-\partial^{\nu} \phi^{\mu}$ with $\phi^{\mu}$ the field of $\phi$-meson. The dual field tensor $\tilde{G}_{\mu \nu}$ is
defined similarly as its electromagnetic counterpart with $F^{\alpha\beta} \rightarrow G^{\alpha\beta}$. Notice that we could
have chosen to describe the $\gamma N N^*$ in the same way as we
describe the $\phi N N^*$ interactions. However, the term proportional
to $g^{(3)}_{\gamma N N^*}$ in  the Lagrangian densities of Eq.
(\ref{phiNNstar}) vanishes in the case of real photon. With the
Lagrangians given in Eqs. (\ref{photonthreehalf}-\ref{phiNNstar}), the
full invariant amplitude of $s$ and $u$ channels can be obtained by following the Feynman rules.

The form factor for the vertices used in the $s$- and $u$-channel
diagrams, $F_{N^*}(p^2)$, is
\begin{equation}
F_{N^*}(p^2)=\frac{\Lambda^{4}}{\Lambda^{4} + (p^2-M^2_{N^*})^2},
\end{equation}
with $\Lambda$ is the cut-off parameter for the virtual $N^*$, following Ref.~\cite{Hung01}. In this work, we choose $\Lambda = 1.2$ GeV for all resonances.
The Rarita-Schwinger propagator is used for the spin-$3/2$ $N^*$
\begin{eqnarray}
G_{\mu\nu}^{(3/2)}(p)&=&\frac{i(\not\! p+M_{N^*})}{p^2-M^2_{N^*}+iM_{N^*}\Gamma_{N^*}}\left[-g_{\mu\nu} + {1 \over 3} \gamma_\mu \gamma_\nu \right.\nonumber \\
&-& \left. {1 \over 3 M_{N^*}} \left(p_\mu \gamma_\nu - p_\nu \gamma_\mu\right) + {2 \over 3 M_{N^*}^2} p_\mu p_\nu\right],\nonumber \\
\end{eqnarray}
with $\Gamma_{N^*}$ the total decay width of $N^*$. Because $u<0$, we take $\Gamma_{N^*} = 0 $ MeV for the propagator in the $u$ channel.

Note carefully that in our previous works \cite{Kiswandhi10, Kiswandhi12}, we cannot obtain the values of the coupling constants $g_{\phi p p^*}$ and $g_{\gamma p p^*}$
by fitting to the experimental data, as our calculations are done in the tree level. Thus, only the values of the products of the coupling constants
$G_{\gamma p p^*}^{ij} \equiv g_{\gamma p p^*}^{(i)}g_{\phi p p^*}^{(j)}$ are shown. Here, we will show how we calculate the neutron coupling constants $g_{\phi n n^*}^{(j)}$
and $g_{\gamma n n^*}^{(i)}$.

First of all, we have to realize that we do not have sufficient knowledge to actually estimate the neutron coupling constants $g_{\phi n n^*}^{(j)}$ and
$g_{\gamma n n^*}^{(i)}$ from the experimental data. At best, we can only assume that our resonance, if it actually exists, should have properties rather similar to a
theoretically predicted $J^P = 3/2^-$ nucleon state in the same mass region that has the same sign for the ratio of helicity amplitudes $A^p_{1/2}/A^p_{3/2}$ for
$p^* \to \gamma p$. For this purpose, we employ a $J^P = 3/2^-$ nucleon state with a bare mass of 2095 GeV and a positive value for the ratio of helicity amplitudes
$A^p_{1/2}/A^p_{3/2}$ for $\gamma p$ decay predicted by Ref. \cite{Capstick92} in which the predictions for Breit-frame
helicity amplitudes $A_{1/2}^N$ and $A_{3/2}^N$ for both $\gamma p$ and $\gamma n$ decays are
\beqna
&& A^p_{1/2} = -9 ;\qquad A^p_{3/2} = -14 \nonumber \\
&& A^n_{1/2} = 8 ;\qquad A^n_{3/2} = 1,
\label{helicity_A}
\eeqna
in the unit of $10^{-3}~\textrm{GeV}^{-1/2}$. Notice that these helicity amplitudes are calculated in Breit frame, not in the center-of-mass frame.
It is also very interesting to note that this predicted state is the only nucleon state with $J^P = 3/2^-$ and a positive ratio of
helicity amplitudes $A_{1/2}^p/A_{3/2}^p$ in the energy region.

Let us also state the relation between the helicity and invariant amplitudes in the center-of-mass frame
of the resonance $N^*$
\beqn
A_m^N = \frac{1}{\sqrt{2 |{\mathbf k}|}} \frac{1}{\sqrt{2 M_{N^*}}} \frac{1}{\sqrt{2 M_N}} {\mathcal M}_{N^*\to \gamma N}(m, m_\gamma = +1)
\eeqn
where $\mathbf k$ is the three-momentum of the photon in center-of-mass frame and $m = 1/2, \ldots, J$ is the spin projection of the resonance in which $J$ is the total
spin of the resonance. The photon is assumed to be moving to the positive $z$ direction with spin projection $m_\gamma = +1$.

By using the values for the helicity amplitudes $A^n_{1/2}$ and $A^n_{3/2}$ for the $n^* \to \gamma n$, it is straightforward to calculate the coupling constants
$g_{\gamma n n^*}$. First, let us translate the coupling constants into helicity amplitudes by using a linear transformation, since they are related only linearly,
\beqn
\left(
\barr{c}
A_{1/2}^{N} \\
A_{3/2}^{N}
\earr
\right)=
\left(
\barr{cc}
\Lambda_{11} & \Lambda_{12} \\
\Lambda_{21} & \Lambda_{22}
\earr
\right)
\left(
\barr{c}
g_{\gamma N N^*}^{(1)} \\
g_{\gamma N N^*}^{(2)}
\earr
\right),
\label{g_to_A}
\eeqn
where
\beqn
\Lambda_{ij} = A_{i-1/2}^N\Big|_{g_{\gamma N N^*}^{(j)} = 1, g_{\gamma N N^*}^{(\ne j)} = 0}
\eeqn
where $m$ and $m_\gamma$ are the spin projections of the resonance $N^*$ and the photon, respectively. Here, $N = (p, n)$ and $N^* = (p^*, n^*)$.
It must also be noted that since the helicity amplitudes are defined for real photon, there is no $m_\gamma = 0$ with $m_N = +1/2$ state for the $A_{1/2}^N$ amplitude.
Finally, in order to obtain $g_{\gamma n n^*}^{(i)}$, we can just inverse the relation
\beqn
\left(
\barr{c}
g_{\gamma n n^*}^{(1)} \\
g_{\gamma n n^*}^{(2)}
\earr
\right) =
\left(
\barr{cc}
\Lambda_{11} & \Lambda_{12} \\
\Lambda_{21} & \Lambda_{22}
\earr
\right)^{-1}
\left(
\barr{c}
A_{1/2}^{n} \\
A_{3/2}^{n}
\earr
\right)
\eeqn

Next, we will calculate the coupling constants $g_{\phi n n^*}^{(j)}$. Notice that by isospin consideration, the values for these coupling constants are the same
for both proton and neutron cases $g_{\phi p p^*}^{(j)} = g_{\phi n n^*}^{(j)}$. First of all, we notice that we only have the products of
$G_{\gamma p p^*}^{ij} \equiv g_{\gamma p p^*}^{(i)} g_{\phi p p^*}^{(j)}$ in our previous work \cite{Kiswandhi12}, and the values of $g_{\gamma p p^*}^{(i)}$ here cannot be
similar to those obtained in Ref. \cite{Capstick92}, as they are obtained in completely different ways. But, we have to find a way to obtain $g_{\gamma p p^*}^{(i)}$
if we would like to get $g_{\phi p p^*}^{(j)}$ from $G_{\gamma p p^*}^{ij}$. However, we can require that the resonance in our work and in Ref. \cite{Capstick92} should have
the same decay width to $\gamma p$ in order to fix $g_{\gamma p p^*}^{(i)}$. Let us also write the partial width of the nucleon resonance $N^*$ decay into $\gamma N$
in terms of the helicity amplitudes
\beqn
\Gamma_{N^* \to \gamma N} = \frac{2}{2J+1} \frac{|{\mathbf k}|^2 M_N}{\pi^2 M_{N^*}} \left(|A_{1/2}^N|^2 + |A_{3/2}^N|^2\right).
\label{width}
\eeqn
Then, the $p^* \to \gamma p$ width in Ref. \cite{Capstick92} is
\beqn
\Gamma_{p^* \to \gamma p} = f_A \left(|A^{p}_{1/2}|^2 + |A^{p}_{3/2}|^2\right)
\eeqn
in which $f_A$, which can be easily read from Eq. (\ref{width}), is just a factor containing kinematical variables. At the same time, as we explain before,
this width has to be equal to the width we have from the coupling constants $g_{\gamma p p^*}^{(i)}$
\beqna
\Gamma_{p^* \to \gamma p} &=& f_g^{(1)} (g_{\gamma p p^*}^{(1)})^2 + f_g^{(2)} (g_{\gamma p p^*}^{(2)})^2 \nonumber \\
&=& f_g^{(1)} (g_{\gamma p p^*}^{(1)})^2 + f_g^{(2)} (k_{12} g_{\gamma p p^*}^{(1)})^2 \nonumber \\
&=& \left(f_g^{(1)} + f_g^{(2)} k_{12}^2\right) (g_{\gamma p p^*}^{(1)})^2
\eeqna
where $k_{12} \equiv g_{\gamma p p^*}^{(2)}/ g_{\gamma p p^*}^{(1)} = G_{\gamma p p^*}^{2j}/G_{\gamma p p^*}^{1j}$ and $f_g^{(i)}$ is just a factor containing kinematical
variables. This leads us to
\beqna
g_{\gamma p p^*}^{(1)} &=& \pm \sqrt{\frac{f_A}{f_g^{(1)} + f_g^{(2)} k_{12}^2}} \sqrt{|A^{p}_{1/2}|^2 + |A^{p}_{3/2}|^2} \nonumber \\
g_{\gamma p p^*}^{(2)} &=& \pm k_{12}\sqrt{\frac{f_A}{f_g^{(1)} + f_g^{(2)} k_{12}^2}} \sqrt{|A^{p}_{1/2}|^2 + |A^{p}_{3/2}|^2} \nonumber \\
\eeqna
which can be written compactly as
\beqna
g_{\gamma p p^*}^{(i)} &=& \pm k_{1i}\sqrt{\frac{f_A}{f_g^{(1)} + f_g^{(2)} k_{12}^2}} \sqrt{|A^{p}_{1/2}|^2 + |A^{p}_{3/2}|^2} \nonumber \\
\eeqna
where $k_{11} = 1$. Notice that the ambiguity in the sign can be resolved easily by substituting the coupling constants to Eq. (\ref{g_to_A}) and requiring the resulting
helicity amplitudes to have the same sign as the ones given in Ref. \cite{Capstick92}.
Once we obtain $g_{\gamma p p^*}^{(i)}$, we can also obtain $g_{\phi p p^*}^{(j)} = g_{\phi N N^*}^{(j)}$
\beqna
g_{\phi N N^*}^{(j)} &=& \frac{G_{\gamma p p^*}^{ij}}{g_{\gamma p p^*}^{(i)}} = \sqrt{\frac{f_g^{(1)} + f_g^{(2)} k_{12}^2}{f_A}} \nonumber \\
&&\times \frac{G_{\gamma p p^*}^{ij}}{k_{1i}\sqrt{|A^{p}_{1/2}|^2 + |A^{p}_{3/2}|^2}}.
\eeqna

\section{Spin-density matrix formalism for $\gamma d \to \phi pn$ reaction}

Let us begin by defining the SDME from the decay angular distribution of the $K \bar{K}$ pair in an arbitrary frame for a specific $t$ and photon polarization state
\cite{schillingnpb15397}
$|\gamma\rangle$
\beqna
W(\Omega_{K\bar{K}})\Big|_{t_\phi} &=& \frac{1}{N} \int d u_\phi d \Omega_{pn}\frac{p_{pn}}{M_{pn}} \nonumber \\
&&\times\sum_{\lambda_p \lambda_n \lambda_d} |\langle \Omega_{K\bar{K}} \lambda_p \lambda_n \big|\hat{T}\big|\gamma \lambda_d\rangle|^2
\label{W_def_d}
\eeqna
with a normalization factor
\beqna
N &=& \frac{1}{2} \int d \Omega_{K\bar{K}} \int d u_\phi d \Omega_{pn} \frac{p_{pn}}{M_{pn}} \nonumber \\
&&\times\sum_{\lambda_p \lambda_n \lambda_\gamma \lambda_d} \big|\langle \Omega_{K\bar{K}} \lambda_p \lambda_n \big|\hat{T}\big|\lambda_\gamma \lambda_d\rangle\big|^2,
\eeqna
where all the kinematic variables have been defined before in Sec. \ref{sec:model}, except the solid angle $\Omega_{K\bar{K}}$ which is the direction of either
$K$ or $\bar{K}$ in $K \bar{K}$ pair rest frame. Here, $\lambda_i$ denotes the helicity of particle $i$. Notice that the integrations over $u$ and $\Omega_{pn}$ are needed
since we do not observe these variables in the final state. The normalization factor $N$ is proportional with the unpolarized DCS, with a factor of $1/2$ needed to average
over the helicity states of the photon and an integration over $\Omega_{K \bar{K}}$ is also included. Clearly, the integration has to be done in the same way we would
integrate for DCS since it is proportional to the decay angular distribution. Here, we have dropped all the momenta, and it is understood that
the momenta of the particles have been fixed by the total energy squared $s$, $t_\phi$, $u_\phi$, $\Omega_{pn}$, and $\Omega_{K \bar{K}}$ together with the use of a coordinate system.

At this point, let us now introduce the photon spin-density operator
\beqn
\hat{\rho}(\gamma) =  |\gamma\rangle\langle \gamma|,
\eeqn
which can be represented by a matrix whose elements are
\beqn
\rho_{\lambda_\gamma \lambda_\gamma'}(\gamma) = \langle \lambda_\gamma |\hat{\rho}(\gamma)| \lambda_\gamma' \rangle.
\eeqn

Then, we can write Eq. (\ref{W_def_d}) as
\beqna
&&W(\Omega_{K\bar{K}})\Big|_{t_\phi} = \frac{1}{N} \int d u_\phi d \Omega_{pn} \frac{p_{pn}}{M_{pn}} \sum_{\lambda_p \lambda_n \lambda_d \lambda_\gamma \lambda_\gamma'}
\nonumber \\
&&\langle \Omega_{K\bar{K}} \lambda_p \lambda_n \big|\hat{T}\big|\lambda_\gamma \lambda_d\rangle \rho_{\lambda_\gamma \lambda_\gamma'}(\gamma)\langle \lambda_\gamma' \lambda_d\big|\hat{T}^\dagger\big|\Omega_{K\bar{K}} \lambda_p \lambda_n \rangle, \nonumber \\
\label{W_def_d_1}
\eeqna
with a normalization factor, written in a different way,
\beqna
N &=&  \int d \Omega_{K\bar{K}} \int d u_\phi d \Omega_{pn} \frac{p_{pn}}{M_{pn}} \sum_{\lambda_p \lambda_n \lambda_d \lambda_\gamma \lambda_\gamma'}
\nonumber \\
&&\times \langle \Omega_{K\bar{K}} \lambda_p \lambda_n \big|\hat{T}\big|\lambda_\gamma \lambda_d\rangle \left(\frac{1}{2} \delta_{\lambda_\gamma \lambda_\gamma'} \right)
\nonumber \\
&&\times \langle \lambda_\gamma' \lambda_d\big|\hat{T}^\dagger\big|\Omega_{K\bar{K}} \lambda_p \lambda_n \rangle,
\eeqna
which shows its similarity with the numerator, with an integration over
$\Omega_{K\bar{K}}$ and $\rho_{\lambda_\gamma \lambda_\gamma'}(\gamma) \to 1/2 \delta_{\lambda_\gamma \lambda_\gamma'}$ where $\delta_{\lambda_\gamma \lambda_\gamma'}$ here
is the Kronecker delta.

Here, by following the same construction, we can write the decay angular distribution as a function of the $\phi$ meson SDME,
\beqna
&& W(\Omega_{K\bar{K}})\Big|_{t_\phi} = \frac{1}{N_\phi} \nonumber \\
&& \sum_{\lambda_\phi \lambda_\phi'} \langle \Omega_{K\bar{K}}\big|\hat{T}_\phi \big|\lambda_\phi\rangle \rho_{\lambda_\phi \lambda_\phi'}(\phi)
\langle \lambda_\phi' \big| \hat{T}_\phi^\dagger \big| \Omega_{K\bar{K}} \rangle,
\label{W_2_d}
\eeqna
where the normalization constant is
\beqna
N_\phi &=& \int d \Omega_{K\bar{K}} \sum_{\lambda_\phi \lambda_\phi'} \langle \Omega_{K\bar{K}}\big|\hat{T}_\phi \big|\lambda_\phi\rangle
\left(\frac{1}{3} \delta_{\lambda_\phi \lambda_\phi'} \right) \nonumber \\
&& \times \langle \lambda_\phi' \big| \hat{T}_\phi^\dagger \big| \Omega_{K\bar{K}} \rangle \nonumber \\
&=& \frac{1}{3} \sum_{\lambda_\phi} \int d \Omega_{K\bar{K}} |\langle \Omega_{K\bar{K}}\big|\hat{T}_\phi \big|\lambda_\phi\rangle|^2,
\label{norm_W_2_d}
\eeqna
where the factor of $1/3$ comes from the averaging of the polarization of the $\phi$ meson.
Now, our main task is to isolate $\rho_{\lambda_\phi \lambda_\phi'}(\phi)$ from Eq. (\ref{W_def_d_1}). We can begin by noticing that
\beqna
\langle \Omega_{K\bar{K}} \lambda_p \lambda_n \big|\hat{T}\big|\lambda_\gamma \lambda_d\rangle &=& c\sum_{\lambda_\phi}
\langle \Omega_{K\bar{K}}\big|\hat{T}_\phi\big|\lambda_\phi \rangle \nonumber \\
&& \times \langle \lambda_\phi \lambda_p \lambda_n \big|\hat{T}_{\gamma d}\big|\lambda_\gamma \lambda_d\rangle,
\eeqna
where a complex number $c$ is produced as we take only the pole of the $\phi$-meson propagator, which is a good approximation as $\phi$ meson has very small width.

After some arranging of the terms, Eq. (\ref{W_def_d_1}) can now be written as
\beqna
&& W(\Omega_{K\bar{K}})\Big|_{t_\phi} = \frac{1}{N} \sum_{\lambda_\phi \lambda_\phi'} \langle \Omega_{K\bar{K}}\big|\hat{T}_\phi \big|\lambda_\phi\rangle
\Bigg[ |c|^2 \sum_{\lambda_p \lambda_n \lambda_d} \sum_{\lambda_\gamma \lambda_\gamma'} \nonumber \\
&& \times \int d u_\phi d \Omega_{pn} \frac{p_{pn}}{M_{pn}}
\langle \lambda_\phi \lambda_p \lambda_n \big|\hat{T}_{\gamma d}\big|\lambda_\gamma \lambda_d\rangle \rho_{\lambda_\gamma \lambda_\gamma'}(\gamma) \nonumber \\
&& \times \langle \lambda_\gamma' \lambda_d\big|\hat{T}_{\gamma d}^\dagger\big|\lambda_\phi' \lambda_p \lambda_n \rangle \Bigg]
\langle \lambda_\phi' \big| \hat{T}_\phi^\dagger \big| \Omega_{K\bar{K}} \rangle,
\label{W_d}
\eeqna
Now, let us evaluate the normalization constant
\beqna
N &=& \frac{1}{2} \int d \Omega_{K\bar{K}} \int d u_\phi d \Omega_{pn} \frac{p_{pn}}{M_{pn}} \nonumber \\
&& \times |c|^2 \sum_{\lambda_p \lambda_n \lambda_\gamma \lambda_d} \big|\langle \Omega_{K\bar{K}} \lambda_p \lambda_n \big|\hat{T}\big|\lambda_\gamma \lambda_d\rangle\big|^2 \nonumber \\
&=& \frac{1}{2}\int d u_\phi d \Omega_{pn} \frac{p_{pn}}{M_{pn}} |c|^2 \sum_{\lambda_\phi \lambda_p \lambda_n \lambda_\gamma \lambda_d} \int d \Omega_{K\bar{K}}\nonumber \\
&& \times \big|\langle \Omega_{K\bar{K}}\big|\hat{T}_\phi\big|\lambda_\phi \rangle \big|^2
\big|\langle \lambda_\phi \lambda_p \lambda_n \big|\hat{T}_{\gamma d}\big|\lambda_\gamma \lambda_d\rangle\big|^2.
\label{W_norm_d}
\eeqna
We can now use the fact that the the $K$ meson is a spinless particle, which allows us to write
\beqn
\langle \Omega_{K\bar{K}}\big|\hat{T}_\phi\big|\lambda_\phi \rangle = A \sqrt{\frac{3}{4 \pi}} D_{\lambda_\phi 0}^1(\Omega_{K\bar{K}})
\eeqn
and
\beqn
\int d \Omega_{K\bar{K}} \big|\langle \Omega_{K\bar{K}}\big|\hat{T}_\phi\big|\lambda_\phi \rangle \big|^2 = |A|^2.
\eeqn
We can factor out $\int d \Omega_{K\bar{K}} \big|\langle \Omega_{K\bar{K}}\big|\hat{T}_\phi\big|\lambda_\phi \rangle \big|^2$ since it is independent of the value of
$\lambda_\phi$ and obtain
\beqna
N &=& \frac{1}{3}\sum_{\lambda_\phi'} \int d \Omega_{K\bar{K}} \big|\langle \Omega_{K\bar{K}}\big|\hat{T}_\phi\big|\lambda_\phi' \rangle \big|^2 \nonumber \\
&& \times \Bigg[\frac{1}{2} \int d u_\phi d \Omega_{pn} \frac{p_{pn}}{M_{pn}} \nonumber
\eeqna
\beqna
&& \times |c|^2 \sum_{\lambda_\phi \lambda_p \lambda_n \lambda_\gamma \lambda_d} \big|\langle \lambda_\phi \lambda_p \lambda_n \big|\hat{T}_{\gamma d}\big|\lambda_\gamma \lambda_d\rangle\big|^2\Bigg],\nonumber \\
\label{norm_d}
\eeqna
where the summation over $\phi$-meson helicity states $\lambda_\phi'$ and a factor of $1/3$ are needed for comparison with Eq. (\ref{norm_W_2_d})

We can compare Eqs. (\ref{W_d}) and (\ref{norm_d}) to Eqs. (\ref{W_2_d}) and (\ref{norm_W_2_d}) to obtain
\beqna
\rho_{\lambda_\phi \lambda_\phi'}(\phi) &\equiv& \frac{1}{N_{\gamma d}} \sum_{\lambda_p \lambda_n \lambda_d} \sum_{\lambda_\gamma \lambda_\gamma'} \int d u_\phi d \Omega_{pn}
\frac{p_{pn}}{M_{pn}} \nonumber \\
&& \times \langle \lambda_\phi \lambda_p \lambda_n \big|\hat{T}_{\gamma d}\big|\lambda_\gamma \lambda_d\rangle
\rho_{\lambda_\gamma \lambda_\gamma'}(\gamma) \nonumber \\
&& \times \langle \lambda_\gamma' \lambda_d\big|\hat{T}_{\gamma d}^\dagger\big|\lambda_\phi' \lambda_p \lambda_n \rangle,
\eeqna
with
\beqna
N_{\gamma d} &\equiv& \frac{1}{2} \int d u_\phi d \Omega_{pn} \frac{p_{pn}}{M_{pn}} \nonumber \\
&& \times\sum_{\lambda_\phi \lambda_p \lambda_n \lambda_\gamma \lambda_d} \big|\langle \lambda_\phi \lambda_p \lambda_n \big|\hat{T}_{\gamma d}\big|\lambda_\gamma \lambda_d\rangle\big|^2.
\eeqna

\section{Relation between single-scattering DCS and the DCS with on-shell $pn$ FSI}
\label{app:DCS_relation}

Let us begin from the operator of the $\gamma d \to \phi p n$ reaction,
\beqna
\hat{T} &=& \hat{T}_S + \hat{T}_{{\textsl{FSI}}},
\eeqna
where the operator $\hat{T}_S$ contains only single-scattering interactions, as represented in Fig. \ref{all_diagram}(a), and
\beqn
-i\hat{T}_{{\textsl{FSI}}} = (-i\hat{T}_{pn})\frac{i}{E - \hat{H} + i\epsilon} (-i\hat{T}_S),
\label{FSI}
\eeqn
in which the operator $\hat{T}_{{\textsl{FSI}}}$ contains $pn$ FSI only, as represented in Fig. \ref{all_diagram}(b).
Here, the total energy $E$ of the final $\phi$, $p$, and $n$ is $E = E_{pn} + E_\phi$, where the energy of the proton and neutron in the final state $E_{pn} = E_p + E_n$.
The free hamiltonian of the particles in the intermediate state where proton, neutron, and $\phi$ meson are not interacting is
\beqna
\hat{H} &=& \hat{H}_{pn} + \hat{H}_\phi
\eeqna
where $\hat{H}_{pn}$ is just the free hamiltonian of the $pn$ system and $\hat{H}_\phi \to E_\phi$ as the $\phi$ meson is already in free space.
Now, since
\beqn
\frac{1}{E_{pn} - \hat{H}_{pn} + i\epsilon} = \mathcal{P}\frac{1}{E_{pn} - \hat{H}_{pn}} - i\pi\delta(\Delta E_{pn}),
\eeqn
Eq. (\ref{FSI}) can be rewritten as
\beqn
\hat{T}_{{\textsl{FSI}}} = \hat{T}_{{\textsl{FSI, off}}} + \hat{T}_{{\textsl{FSI, on}}}.
\eeqn
However, let us focus on the amplitude calculated within the on-shell approximation, or
\beqn
\hat{T}_{{\textsl{on}}} = \hat{T}_S + \hat{T}_{{\textsl{FSI, on}}}
\eeqn
where again the subscript ${\textsl{on}}$ on the full operator $\hat{T}_{{\textsl{on}}}$ is to denote that the on-energy-shell approximation has been taken, and
in which
\beqn
\hat{T}_{{\textsl{FSI, on}}} \equiv \hat{T}_{pn} \left[-i\pi\delta(\Delta E_{pn})\right] \hat{T}_S.
\eeqn
where the operator $\hat{T}_{pn}$ is defined for elastic $pn$ scattering in which the initial and final states are both on energy shell and
$\Delta E_{pn}$ is the difference of the on-shell energies between the two states.

Here, we need to have some basic relations between the amplitudes involved in $pn$ scattering.
The scattering operator
\beqn
\hat{S}_{pn} = \hat{1} + 2\pi\delta(\Delta E_{pn})(-i\hat{T}_{pn}),
\label{S}
\eeqn
obeys the unitarity relation
\beqn
\hat{S}_{pn} \hat{S}_{pn}^\dagger = \hat{S}_{pn}^\dagger \hat{S}_{pn} = \hat{1},
\label{unitary}
\eeqn
which also implies that
\beqn
\hat{T}_{pn}\hat{T}_{pn}^\dagger = \hat{T}_{pn}^\dagger\hat{T}_{pn}.
\eeqn

Having all these relations, we can continue to calculate the operator $\hat{T}_{{\textsl{on}}}$
\beqna
\hat{T}_{{\textsl{on}}} &=& \hat{T}_S + \hat{T}_{pn} \left[- i\pi\delta(\Delta E_{pn})\right] \hat{T}_S \nonumber \\
&=& \frac{1}{2} \left(\hat{1} + \hat{S}_{pn}\right)\hat{T}_S.
\eeqna
Now, we are ready to calculate $\hat{T}_{{\textsl{on}}}^\dagger \hat{T}_{{\textsl{on}}}$
\beqna
\hat{T}_{{\textsl{on}}}^\dagger \hat{T}_{{\textsl{on}}} &=& \frac{1}{4} \hat{T}_S^\dagger \left(\hat{1} + \hat{S}_{pn}^\dagger\right) \left(\hat{1} + \hat{S}_{pn}\right)\hat{T}_S \nonumber \\
&=& \frac{1}{4} \hat{T}_S^\dagger \left[4 \hat{1} - 2i\pi\delta(\Delta E_{pn})\hat{T}_{pn} \right. \nonumber \\
&& \left. + 2i\pi\delta(\Delta E_{pn})\hat{T}_{pn}^\dagger \right] \hat{T}_S,
\eeqna
where the last line is reached by using Eqs. (\ref{S}, \ref{unitary}). We can continue
\beqna
\hat{T}_{{\textsl{on}}}^\dagger \hat{T}_{{\textsl{on}}}
&=& \hat{T}_S^\dagger \left[\hat{1} - \frac{1}{2} i\pi\delta(\Delta E_{pn})\hat{T}_{pn} \right. \nonumber \\
&& \left. + \frac{1}{2} i\pi\delta(\Delta E_{pn})\hat{T}_{pn}^\dagger \right] \hat{T}_S \nonumber \\
&=& \hat{T}_S^\dagger \hat{T}_S + \frac{1}{2} \hat{T}_S^\dagger \hat{T}_{pn} \left[-i\pi\delta(\Delta E_{pn})\right] \hat{T}_S \nonumber \\
&& + \frac{1}{2} \hat{T}_S^\dagger \left[i\pi\delta(\Delta E_{pn})\right] \hat{T}_{pn}^\dagger \hat{T}_S \nonumber \\
&=& \hat{T}_S^\dagger \hat{T}_S + \frac{1}{2} \left(\hat{T}_S^\dagger \hat{T}_{{\textsl{FSI, on}}} + \hat{T}_{{\textsl{FSI, on}}}^\dagger \hat{T}_S\right)
\label{T_on_calc}
\eeqna
Finally, by using
\beqna
\hat{T}_{{\textsl{on}}}^\dagger \hat{T}_{{\textsl{on}}} &=& \left(\hat{T}_S^\dagger + \hat{T}_{{\textsl{FSI, on}}}^\dagger \right) \left(\hat{T}_S + \hat{T}_{{\textsl{FSI, on}}}\right)\nonumber \\
&=& \hat{T}_S^\dagger \hat{T}_S + \hat{T}_{{\textsl{FSI, on}}}^\dagger \hat{T}_{{\textsl{FSI, on}}} + \hat{T}_S^\dagger \hat{T}_{{\textsl{FSI, on}}} \nonumber \\
&& + \hat{T}_{{\textsl{FSI, on}}}^\dagger \hat{T}_S
\eeqna
to substitute the second term in the round brackets of the last line of Eq. (\ref{T_on_calc}), we have, from Eq. (\ref{T_on_calc})
\beqn
\hat{T}_{{\textsl{on}}}^\dagger \hat{T}_{{\textsl{on}}} = \hat{T}_S^\dagger \hat{T}_S - \hat{T}_{{\textsl{FSI, on}}}^\dagger \hat{T}_{{\textsl{FSI, on}}}.
\label{T_resc}
\eeqn
Now, we must be very careful in interpreting this apparently simple formula. It has to be understood that this equation consists of operators and can be related to
amplitudes only after we apply a suitable set of states onto them. This procedure also includes an application of a completeness relation between the operators.
The completeness relation in this case must be constructed from the states present in the $S$-matrix $\hat{S}_{pn}$, which when applied
to the operators provides a summation over the spins and an integration over the solid angles of the proton and neutron. The resulting quantities will be proportional
to the DCS of incoherent $\phi$ meson photoproduction in which the momenta and spins of the outgoing proton and neutron is not observed,
\beqn
\frac{d \sigma_{{\textsl{total, on}}}}{dt} = \frac{d \sigma_S}{dt} - \frac{d \sigma_{{\textsl{FSI, on}}}}{dt},
\eeqn
which proves the statement stated before. Notice that unitarity relation for the $pn$ system in the final state given in Eq. (\ref{unitary}) actually plays an important
role in the derivation.

\end{document}